\documentclass[a4paper,fleqn,usenatbib]{mnras}

\usepackage{times}

\usepackage[T1]{fontenc}
\usepackage{ae,aecompl}

\usepackage{enumitem}

\usepackage{graphicx}	
\usepackage{amsmath}		
\usepackage{amssymb}		

\newcommand{\revised}{\rm}

\title[Scattered light images of spiral arms in discs]{Scattered light images of spiral arms in marginally gravitationally unstable discs with an embedded planet}

\author[A. Pohl et al.]
{
A. Pohl$^{1,2}$\thanks{E-mail:pohl@mpia.de},
P. Pinilla$^{3}$, M. Benisty$^{4}$, S. Ataiee$^{5,6}$, A. Juh\'asz$^{7}$, C.P. Dullemond$^{2}$,\newauthor R. Van Boekel$^{1}$ and T. Henning$^{1}$\\
$^{1}$Max-Planck-Institute for Astronomy, K\"onigstuhl 17, D-69117 Heidelberg, Germany\\
$^{2}$Heidelberg University, Institute of Theoretical Astrophysics, Albert-Ueberle-Str. 2, D-69120 Heidelberg, Germany\\
$^{3}$Leiden Observatory, Leiden University, P.O. Box 9513, NL-2300 RA Leiden, The Netherlands\\
$^{4}$University Grenoble Alpes, IPAG, F-38000 Grenoble, France; CNRS, IPAG, F-38000 Grenoble, France\\
$^{5}$School of Astronomy, Institute for Research in Fundamental Sciences (IPM), Tehran, Iran\\
$^{6}$Physics Institute, Space Research and Planetary Sciences, Sidlerstrasse 5, CH-3012 Bern, Switzerland\\
$^{7}$Institute of Astronomy, Madingley Road, Cambridge CB3 OHA, United Kingdom\\
}

\date{Accepted 2015 July 28.}

\pubyear{2015}

\begin{document}
\label{firstpage}
\pagerange{\pageref{firstpage}--\pageref{lastpage}}
\maketitle

\begin{abstract}
Scattered light images of transition discs in the near-infrared often show non-axisymmetric structures in the form of wide-open spiral arms in addition to their characteristic low-opacity inner gap region. We study self-gravitating discs and investigate the influence of gravitational instability on the shape and contrast of spiral arms induced by planet-disc interactions. Two-dimensional non-isothermal hydrodynamical simulations including viscous heating and a cooling prescription are combined with three-dimensional dust continuum radiative transfer models for direct comparison to observations. We find that the resulting contrast between the spirals and the surrounding disc in scattered light is by far higher for pressure scale height variations, i.e. thermal perturbations, than for pure surface density variations. Self-gravity effects suppress any vortex modes and tend to reduce the opening angle of planet-induced spirals, making them more tightly wound. If the disc is only marginally gravitationally stable with a Toomre parameter around unity, an embedded massive planet (planet-to-star mass ratio of $10^{-2}$) can trigger gravitational instability in the outer disc. The spirals created by this instability and the density waves launched by the planet can overlap resulting in large-scale, more open spiral arms in the outer disc. The contrast of these spirals is well above the detection limit of current telescopes.
\end{abstract}




\section{Introduction}
\label{sec:introduction}

Young forming protoplanets may leave observational signatures in their parental disc in the form of distinct gaps, vortices, warps, or spiral arms. This is thought to happen during the final stages of protoplanetary disc evolution. There is a peculiar group called transition discs, whose spectral energy distribution (SED) and sub-millimetre (sub-mm) observations suggest that the inner disc is strongly depleted of dust. Spiral arms have thus far been only observed in this kind of discs, suggesting that the gap formation mechanism is related to spiral arm formation. This needs, however, to be treated carefully since there might be an observational selection effect. Observations in the near-infrared (NIR) and (sub-)mm regime of transition discs show a variety of asymmetric features. At sub-mm wavelengths asymmetries are seen for example in the system Oph IRS 48 \citep{vandermarel2013}, SR 21, HD 135344B/SAO 206462 \citep{perez2014}, and HD 142527 (\citealt{casassus2013, fukagawa2013}), for the latter a low-mass stellar companion has been found (\citealt{biller2012, close2014}). Since discs are optically thick in the optical and NIR, high angular resolution scattered light images trace small hot dust particles in the upper disc layer. Most discs have double spirals with a large opening angle that are almost symmetric, e.g. MWC 758 (\citealt{grady2013,benisty2015}), LkCa~15 (\citealt{thalmann2014}), HD 135344B/SAO 206462 \citep{muto2012,garufi2013}, HD142527 (\citealt{casassus2012,rameau2012,canovas2013,avenhaus2014}) and HD100546 (\citealt{boccaletti2013,avenhaus2014b}).\\

The origin of the observed spirals is still debated. For instance, with a simple analytical description based on the spiral density wave theory, the morphology of spirals triggered by a hypothetical planet has been studied by \citet{muto2012}, \citet{grady2013} and \citet{benisty2015}. The best spiral fitting model requires a high disc aspect ratio $h=H/r$ to account for the large pitch angles (angle between the spiral arm and the tangent circle) implying a very warm disc. Recently, \citet{juhasz2014} focused on the contrast of spirals compared to their surrounding background disc. They modelled the spiral waves launched by planets by means of locally isothermal hydrodynamic simulations as well as analytic descriptions. Based on contrast arguments these authors suggested that the spiral arms observed are the results of pressure scale height perturbations rather than of pure surface density perturbations. \revised{\citet{dong2014} also studied the observational signatures in transition discs by combining two-dimensional two fluid hydrodynamical calculations with three-dimensional radiative transfer simulations, but rather focused on the observational signatures of gaps opened by one or several planets. They stated that density waves and streamers inside the planetary gap, produced by planet-disc interactions, can be visible in NIR images.} Besides processes involving planets, it has been examined whether other mechanisms can lead to spiral features, such as non-ideal magnetohydrodynamical effects (e.g. \citealt{flock2015, lyra2015}) or gravitational instability (e.g. \citealt{lodato2004}, \citeyear{lodato2005}; \citealt{rice2004}). Structures induced by the onset of gravitational instabilities tend to produce spirals with higher than m=2 azimuthal wave number (e.g. \citealt{cossins2009,forgan2011}). Planets drive, depending on the planet-to-star mass ratio, a m=1 or sometimes even a m=2 mode, but the secondary spiral is substantially weaker. Our idea is to combine both spiral formation scenarios in order to see if this strengthens their amplitude and to produce various spiral morphologies.\\

In general, the precondition for a disc to become unstable is that the amount of gravitational potential energy overcomes pressure and rotational kinetic support. The axisymmetric stability of a thin disc is determined by the so-called Toomre parameter $Q$ \citep{toomre1964}. The Toomre criterion for a disc to be unstable is defined as

\begin{equation}
	Q = \frac{c_{\mathrm{s}} \kappa}{\pi G \Sigma_{\mathrm{g}}} \lesssim 1\,,
	\label{eq:toomre}
\end{equation}

\noindent where $c_{\mathrm{s}}$ is the sound speed, $\Sigma_{\mathrm{g}}$ is the gas surface density and $\kappa$ corresponds to the epicyclic frequency, which is equal to the angular velocity $\Omega = \left( G M_{\star}/r^3 \right)^{1/2}$ in the case of a Keplerian disc. Increasing the disc mass and/or decreasing the disc temperature leads to a lower Q. Hence, the disc mass has to be a significant fraction of the stellar mass for the disc to be gravitationally unstable.\\

Current observational methods to estimate the disc mass still have systematic uncertainties. The total disc mass is dominated by the gas, which mostly consists of H$_{2}$. Since its emission is hard to detect, there is only one object (TW Hya) for which hydrogen was directly observed in the form of HD (\citealt{bergin2013}). Two prominent proxies, dust continuum emission and CO and its isotopologues, are usually used to estimate disc gas masses. The first method is based on observations of the mm continuum emission from dust grains (e.g. \citealt{beckwith1990,andrews2005}), assuming a certain dust opacity and gas-to-dust ratio. The generally assumed value of 100 for the gas-to-dust ratio may not be accurate for all discs, since it strongly depends on the disc evolution (e.g. \citealt{brauer2007, birnstiel2010}). Also, from observations it is known that mm-sized dust grains and the gas do not necessarily have the same spatial distribution in discs (e.g. \citealt{degregorio2013, walsh2014}). Furthermore, in this calculation dust opacity values at the observed frequency are assumed, which may have quite uncertain values in protoplanetary discs due to the various grain sizes, compositions, internal structures and the presence of ice mantles (e.g. \citealt{pollack1994,henning1996,semenov2003,demyk2013}). The second method infers the gas mass independently of the dust content using molecular lines observations. Due to its high abundance and line strength CO and its isotopologues are the most frequently used tracers of gas in protoplanetary discs (e.g. \citealt{dutrey1996, williams2014}). Uncertainties concern the H$_{2}$-CO abundance and the topologist ratios because of photo-dissociation and freeze-out processes. \citet{miotello2014} showed that the disc mass may be underestimated by up to two orders of magnitude if only a single CO isotopologue line is observed and isotope selective effects are not properly taken into account. As mentioned above, the only direct measurement of hydrogen in discs from TW Hya actually suggests that the discs might be more massive than previously thought. Therefore, current disc mass estimations from gas and dust observations only give a very rough approximation and discs around Class II stars might be sufficiently massive to be gravitationally unstable, as it is expected for Class 0 and~I objects.\\

All disc processes including planet-disc interactions and gravitational instability strongly depend on the cooling efficiency, and therefore, the disc temperature. The latter again has an important effect on the shape and contrast of non-axisymmetric disc features. However, many of the numerical simulations so far are limited due to the assumption of a locally isothermal disc. In this paper, a two-dimensional hydrodynamical code which includes an energy equation accounting for disc viscous heating and cooling is used to simulate planet-disc interactions. Gravitational instability is known to produce multi-armed spirals in discs, while planet-disc interactions mostly result in one- or two-armed spirals. In this work, we study the role of heating and cooling, and gravitational instability on the shape and contrast of spiral arms induced by planet-disc interactions. We want to investigate what kind of structures the interplay between gravitational instability and planet-disc interactions can reproduce. This combination has not been investigated so far. The question we ask here is whether the presence of a planet in a marginally gravitationally stable disc can tip it over the limit. The information about the hydrodynamical disc structure is a prerequisite for our radiative transfer modelling in order to link simulation results to observations. The focus is on presenting synthetic scattered light images and on investigating whether they resemble the observations. This paper is divided into five parts. In Sect. \ref{sec:hydro_sims} the modifications to the hydrodynamical code and the basic models are introduced. The radiative transfer setup is described in Sect. \ref{sec:rt_modeling}. Subsequently, the simulation results are presented in Sect. \ref{sec:results}. The associated discussion and the conclusions are summarized in Sect. \ref{sec:summary}.

\section{Hydrodynamical setup}
\label{sec:hydro_sims}

The two-dimensional hydrodynamical grid-based code used for simulating the planet-disc interactions is based on the \textsc{fargo-adsg} version \citep{baruteau2008a,baruteau2008b}. This modified version of the original \textsc{fargo} code \citep{masset2000} implements an energy equation and the disc self-gravity. The full treatment of disc heating and cooling is important, since in addition to surface density perturbations, temperature perturbations occur as a consequence of shocks along the spirals.

\subsection{Surface density description}
\label{subsec:hydro_density}
The gas disc is characterized by an initial gas surface density profile $\Sigma_{\mathrm{g}}(r)$, a temperature profile $T(r)$ and a pressure profile $P(r)$. The density profile is taken to be a power law combined with an exponential cut-off at long radii, specifically, 

\begin{equation}
	\Sigma_{\mathrm{g}}(r)=\Sigma_{\mathrm{g,c}} \left( \frac{r}{r_\mathrm{c}} \right)^{-\delta} \exp{}\left[ -\left( \frac{r}{r_{\mathrm{c}}}\right)^{2-\delta}{}\right]\,,
	\label{eq:hydro_lyndenbell}
\end{equation}

\noindent where $r_{\mathrm{c}}$ corresponds to a characteristic scaling radius, which is set to $=75$\,au in accordance with results from high angular resolution disc imaging in the sub-mm regime (\citealt{andrews2010}, \citeyear{andrews2011}). In our simulations this is equal to 0.3 times the outer boundary of the disc. $\Sigma_{\mathrm{g,c}}$ describes the density normalization factor. The surface density index $\delta$ is taken to be 1 in our simulations. With this surface density profile it is ensured that most of the disc mass ($> 60\,\%$) is located between $r_{\mathrm{in}}$ and $r_{\mathrm{c}}$, so that artificial reflections from the outer grid boundary can be minimized.\\

The initial temperature structure is a power law, leading for our case of a non-flaring ($H/r~=~\mathrm{const.}$) disc to 

\begin{equation}
	T(r)=\frac{\mu m_{\mathrm{p}}}{\mathcal{R}_{\mathrm{g}}} G M_{\star} h_{\mathrm{0}}^{2}\,r^{-1}\,,
	\label{eq:bg_temp}
\end{equation}

\noindent with mean molecular weight $\mu$, proton mass $m_{\mathrm{p}}$, universal gas constant $\mathcal{R}_{\mathrm{g}}$, gravitational constant $G$, and where the parameter $h_{\mathrm{0}}$ defines the initial disc aspect ratio.\\

Such a conical disc defines the boundary between a flaring disc, i.e. $H/r$ is a monotonic increasing function of radius r, and a self-shadowed disc, i.e. $H/r$ becomes smaller with larger radii (\citealt{dullemond2004}). Considering that we intend to analyse scattered light images, in the case of our non-flaring, constant opening angle geometry, any pressure scale height perturbation with a sufficient amplitude can cast a shadow over the remaining outer disc rather easily. In this case of grazing incidence an object can cast the largest shadow. Furthermore, the background brightness is as dark as possible for grazing infall. Therefore, high-contrast spiral structures are most easily made in non-flaring discs. Furthermore, \citet{juhasz2014} showed that modelling a flaring disc affects the brightness of the disc in the outer regions, but does not improve the visibility of the spirals. From observations there is also evidence for very low flaring index in MWC 758, one of the sources where a two-armed spiral was detected.

\subsection{Heating and cooling terms}
The development of gravitational instability leads to a self-regulation process due to the competition between heating from the instability and cooling. After a local temperature increase due to a weak shock induced by the spiral density waves, the disc cools back down to a predefined background temperature on a certain time-scale. It takes at least a dynamical time-scale ($t_{\mathrm{dyn}} \sim \Omega^{-1}$) for the disc to hydrodynamically react to any increased temperature. The form of the energy equation implemented in \textsc{fargo} is

\begin{equation}
	\frac{\partial e}{\partial t} + \nabla \cdot (e\,\vec{v}) = -P\nabla \cdot \vec{v} + Q_{+} - Q_{-}\,,
	\label{eq:energy_eq}
\end{equation}
\label{subsec:hydro_cooling}

\noindent where $e$ denotes the thermal energy per unit area, $\vec{v}$ is the flow velocity, $P$ is the vertically integrated pressure, and $Q_{+}/Q_{-}$ corresponds to the vertically integrated heating/cooling term. The disc heating is assumed to be due to the disc viscosity. For the cooling source term we assume

\begin{equation}
	Q_{-} = \frac{e}{t_{\mathrm{cool}}}
	\label{eq:cooling_term}
\end{equation}

\noindent with a cooling time-scale of (cf. \citealt{gammie2001})

\begin{equation}
	t_{\mathrm{cool}} = \beta\,\Omega^{-1}\,,
	\label{eq:cooling_timescale}
\end{equation}

\noindent where the $\beta$-factor is taken to be a constant. The powerlaw background temperature profile to which the disc cools down after each viscous heating event is set by the initial locally isothermal temperature (cf. Eq. \ref{eq:bg_temp}). It is assumed to be due to a balance between viscous heating and irradiation heating on the one hand, and radiative cooling on the other hand. With our choice of the background temperature we expect that local heating by shocks becomes relevant. This shock heating may increase the scale height of the disc, which causes local bumps on the disc surface (radial $\tau=1$ surface with $\tau$ being the optical depth). This is different for a \textit{flaring} irradiated disc ($T(r) \propto r^{-1/2}$) around a Herbig star, where the heating of the disc is dominated by the star apart from the very inner disc. It should be pointed out that in a flaring disc model a different background temperature profile is probably needed.\\

The question is how fast the disc is able to cool down after a shock. If the cooling is very efficient, i.e. $t_{\mathrm{cool}} \ll \Omega^{-1}$, no shock effects will occur and the disc can be assumed to be locally isothermal. For a rough estimation of $t_{\mathrm{cool}}$ in our models, i.e. $\beta$, we do the following calculation. The cooling rate of an accretion disc strongly depends on the optical depth $\tau$, but since we are in the range $\tau \gg 1$, it is determined by

\begin{equation}
	Q_{-}(r) = 2 \sigma T_{\mathrm{eff}}^4(r) = 2 \sigma \frac{1}{\tau} T_{\mathrm{mid}}^4(r)\,,
	\label{eq:coolingrate}
\end{equation}

\noindent where $\sigma$ is the Stefan-Boltzmann constant and T$_{\mathrm{mid}}$ corresponds to the mid-plane temperature. The optical depth can be calculated by

\begin{equation}
	\tau \simeq \frac{1}{2} \Sigma_{\mathrm{d}} \kappa_{\mathrm{d}}\,,
	\label{eq:opticaldepth}
\end{equation}

\noindent whereas a dust-to-gas ratio of 0.01 is assumed and for the dust opacity $\kappa_{\mathrm{d}}$ the Planck mean value is taken. Furthermore, the vertically integrated thermal energy is determined by

\begin{equation}
	E = \frac{1}{\gamma-1} \Sigma_{\mathrm{g}} \frac{k_{\mathrm{B}}T}{\mu m_{\mathrm{p}}}\,,
	\label{eq:thermalener}
\end{equation}

\noindent with Boltzmann constant $k_{\mathrm{B}}$, proton mass $m_{\mathrm{p}}$, adiabatic index $\gamma = \frac{7}{5}$, and mean molecular weight $\mu = 2.3$. Using Eq. \ref{eq:cooling_term} and dividing $t_{\mathrm{cool}}$ by the orbital time-scale $t_{\mathrm{orb}} = 2\,\pi\,\Omega^{-1}$ yields the $\beta$-factor. Using typical values for the gas density ($\Sigma_{\mathrm{g}}(50\,\mathrm{au}) \simeq 50$\,g\,cm$^{-2}$) and temperature ($T_{\mathrm{mid}}(50$\,au) $\simeq\,12$\,K), and setting $\kappa_{\mathrm{d}}(T_{\mathrm{mid}}) \simeq 10$\,cm$^2$g$^{-1}$ gives $\beta \simeq 5$.\\

There is a critical cooling time-scale, $t_{\mathrm{cool,c}}$, and a corresponding critical value of $\beta$, $\beta_{c}$, below which the disc can fragment and form gravitationally bound clumps, rather than reaching a steady, gravitoturbulent state. In numerical experiments \citet{gammie2001} showed that disc fragmentation can happen for a typical time-scale of $t_{\mathrm{cool,c}} \lesssim 3\,\Omega^{-1}$. Using three-dimensional smoothed particle hydrodynamic (SPH) simulations \citet{rice2003} generally confirmed this cooling time for fragmentation, even though the $\beta_{\mathrm{c}}$ value is up to a factor of two higher for more massive discs. However, \citet{paardekooper2012} showed that disc fragmentation is a stochastic process and observed it for cooling times up to $20\,\Omega^{-1}$. In principle, fragmentation is even possible up to $\beta=50$, but it becomes very rare for such high values. \revised{A detailed study on the convergence of the critical cooling time-scale with resolution for SPH and grid-based hydrodynamics simulations was done by \citet{meru2012}. They showed that reducing the dissipation from the numerical viscosity leads to larger values of $\beta_{\mathrm{c}}$ at a given resolution.} In the following we decided to use $\beta$-values of 1 and 10, for which none of our disc models is found to fragment \revised{for the selected resolution}. This allows the disc to settle down into a quasi-steady, self-gravitating state. \revised{We note here that, following the argument from \citet{meru2012}, repeating our simulations with a higher resolution could minimize the artificial viscosity and, therefore, possibly fragment the disc into bound objects.}

\subsection{Parameter choice}
\label{subsec:hydro_setup}
The basic model consists of a viscous, non-flaring disc with an embedded giant planet. An overview of all the parameters can be found in Table \ref{tab_fargo}. For the complete series of models, the disc is tapered as described in Sect. \ref{subsec:hydro_density}. \textsc{fargo} uses dimensionless units, therefore, the fixed orbital radius of the embedded planet of $r_{\mathrm{p}} = 1$ is used as the length scale. The disc is divided into 1280 azimuthal and 720 radial grid zones, ranging from 0.2 to 10.0. This corresponds to a physical disc extension from 5\,au to 250\,au, assuming that the planet's orbital radius is 25\,au. The high resolution for the hydrodynamical part serves to avoid strong numerical diffusion effects. The outer boundary of the computational grid is therefore far enough away from the regions around the gap on which we focus our study. Thus, together with the choice of a tapered gas surface density, artificial boundary condition effects are minimized. The inner boundary conditions are open, allowing for mass outflow at the inner edge. The disc is considered to be non-flaring with a constant aspect ratio of $h=0.05$. \revised{Furthermore, for the disc viscosity the $\alpha$-type viscosity \citep{shakura1973} is used with $\alpha = [10^{-3}, 10^{-2}]$. The planet's potential itself is softened over a length that scales with the disc thickness with a factor of $\epsilon=0.6$. The time-scale over which the planet mass is switched on for the potential evaluation at the beginning of each simulation is set to 100 planetary orbits.} The results of all hydrodynamical simulations are described in Sect. \ref{subsec:results_hydro}.

\section{Radiative transfer setup}
\label{sec:rt_modeling}

The output of the two-dimensional hydrodynamical simulations previously presented in Sect. \ref{sec:hydro_sims}, i.e. the gas surface density distribution $\Sigma_{\mathrm{g}}(r,\varphi)$ and the temperature distribution $T(r,\varphi)$, are taken for further processing in the context of three-dimensional radiative transfer modelling. The radiative transfer code \textsc{radmc-3d}\footnote{http://www.ita.uni-heidelberg.de/$\sim$dullemond/software/radmc-3d/} is used to calculate synthetic scattered light images in the NIR. For these calculations the same aspect ratio, flaring index, and radial and azimuthal grid extensions as in the hydrodynamical simulations are adopted. In order to avoid low photon statistics, the outputs from the hydrodynamical simulations are interpolated to a grid with a lower resolution ($N_{r} \times N_{\varphi} = 384 \times 512$ grid cells) for the radiative transfer calculations. This does not affect the shape of any of the structures obtained in the \textsc{fargo} simulations. In addition, $N_{\theta}$ = 128 cells for the polar direction are taken. Assuming the vertical density profile to be Gaussian and adopting the canonical dust-to-gas ratio for the interstellar medium of 0.01, the dust volume density is given by

\begin{equation}
	\rho(R,\varphi,z) = 0.01\,\frac{\Sigma_{\mathrm{g}}(R,\varphi)}{\sqrt{2\,\pi}\,H(R,\varphi)}\,\exp \left( -\frac{z^2}{2\,H^2(R,\varphi)} \right)\,,
	\label{eq:volume_density}
\end{equation}

\noindent where the spherical coordinates $R$ and $z$ can be converted into cylindrical formulas via $R = r\,\sin(\theta)$ and $z = r\,\cos(\theta)$, where $\theta$ is the polar angle. The temperature output after a specific orbital evolution from the hydrodynamical simulations is used to determine the pressure scale height $H(R,\varphi)$,

\begin{equation}
	H(R,\varphi) = \frac{c_s}{\Omega} = {\sqrt{\frac{k_B}{\mu\,m_{\mathrm{p}}\,G\,M_{\star}}\,T(R,\varphi)\,R^3}} \propto \sqrt{T}\,.
	\label{eq:scale_height}
\end{equation}

The radiative transfer calculations start with computing the dust temperature structure by means of a thermal Monte Carlo simulation using $10^{7}$ photon packages. We want to calculate an equilibrium dust temperature considering the star as the source of luminosity. This temperature is the result of a balance between radiative absorption and re-emission, i.e. a dust grain acquires as much energy as it radiates. This implies that the gas temperature from the hydrodynamical simulations is assumed to only influence the disc scale height (cf. Eq. \ref{eq:scale_height}), but is not explicitly related to the dust temperature itself. Hence, a self-enhancement effect, such that the further irradiation of a disc bump actually influences the scale height at those positions, is not included in our study. The main inputs for the radiative transfer modelling are the dust density structure from Eq. \ref{eq:volume_density}, dust opacities, and the radiation source. The latter is assumed to have typical stellar parameters of a Herbig Ae star ($T_{\mathrm{eff}}=9500\,$K, $M_{\star}=2.0\,$M$_{\odot}$, $R_{\star}=2.5\,$R$_{\odot}$). For simplicity, a black body radiation field is assumed. The dust in our models consists of silicate grains with a fixed grain radius of 0.1\,$\mu$m. The optical constants are obtained from the Jena database based on the work by \citet{jaeger1994} and \citet{dorschner1995}. Scattering is considered to be anisotropic with a full treatment of polarization. The scattering phase function depends on the scattering angle and on the polarization state of the input radiation, i.e. the parameters of the full Stokes vector [I,Q,U,V]. Linearized polarized intensity images can be calculated with $\mathrm{PI} = \left( \mathrm{Q}^2+\mathrm{U}^2 \right)^{1/2}$. The dust opacity tables are produced with the \textsc{BHMIE} code of \citet{bohren1984} for calculating scattering and absorption by spheres. The images are calculated at 1.65\,$\mu$m using $10^{8}$ photon packages.\\

The signal-to-noise ratio of the theoretical images obtained from \textsc{radmc-3d} is limited only by photon statistics. In order to simulate synthetic observations from these models that are comparable to observational data, the images have to be convolved with the telescope's point spread function (PSF). The diffraction limited PSF for a perfect optical system based on circular elements would be an Airy pattern, whose central peak can be approximated by a two-dimensional Gauss function. As size of the PSF a full width at half maximum (FWHM) of 0\farcs04 is chosen, which is representative for \textit{H}-band observations with the SPHERE instrument at the Very Large Telescope (VLT).

\begin{table}
\begin{minipage}{80mm}
	\caption[Overview of the models and parameters we studied.]{Overview of the models and parameters we studied.}
	\label{tab_fargo}
	\begin{tabular}[h]{ccccccc}
		\hline\hline \# & M$_{\mathrm{disc}}$[M$_{\star}$] & $\alpha$ & $M_{\mathrm{p}}/M_{\star}$ & $\beta$ & $r_{c}$ [r$_{\mathrm{p}}$] & SG\\
		\hline
		1 & 0.15 & $10^{-3}$ & $10^{-3}$ & 10 & 3.0 & no \\
		2 & 0.15 & $10^{-3}$ & $10^{-3}$ & 10 & 3.0 & yes \\
		3 & 0.15 & $10^{-3}$ & $10^{-3}$ & 1 & 3.0 & yes \\
		4 & 0.15 & $10^{-2}$ & $10^{-3}$ & 10 & 3.0 & yes \\
		5 & 0.15 & $10^{-3}$ & $10^{-2}$ & 10 & 3.0 & yes \\
		6 & 0.2 & $10^{-3}$ & $10^{-3}$ & 10 & 3.0 & no \\
		7 & 0.2 & $10^{-3}$ & $10^{-3}$ & 10 & 3.0 & yes \\
		8 & 0.2 & $10^{-3}$ & $10^{-2}$ & 10 & 3.0 & yes \\
		9 & 0.2 & $10^{-3}$ & $10^{-3}$ & 10 & 4.8 & yes \\
		\hline  
	\end{tabular}
		
	\medskip
	\textbf{Notes.} \small{When specified the parameters are given in dimensionless \textsc{fargo} units, i.e. with respect to the planet's position and star mass. SG stands for self-gravity. For all simulations $N_{r} \times N_{\varphi} = 720 \times 1280$ grid cells, a radial range from $r_{\mathrm{in}}=0.2\,r_{\mathrm{p}}$ to $r_{\mathrm{out}}=10.0\,r_{\mathrm{p}}$, and open boundary conditions at the inner edge are used. Model~\textit{1} is taken as the reference model considering a non self-gravitating disc.}
\end{minipage}
\end{table}

\section{Results}
\label{sec:results}

\begin{figure*}
	\centering
	\centerline{
		\includegraphics[width=1.0\textwidth]{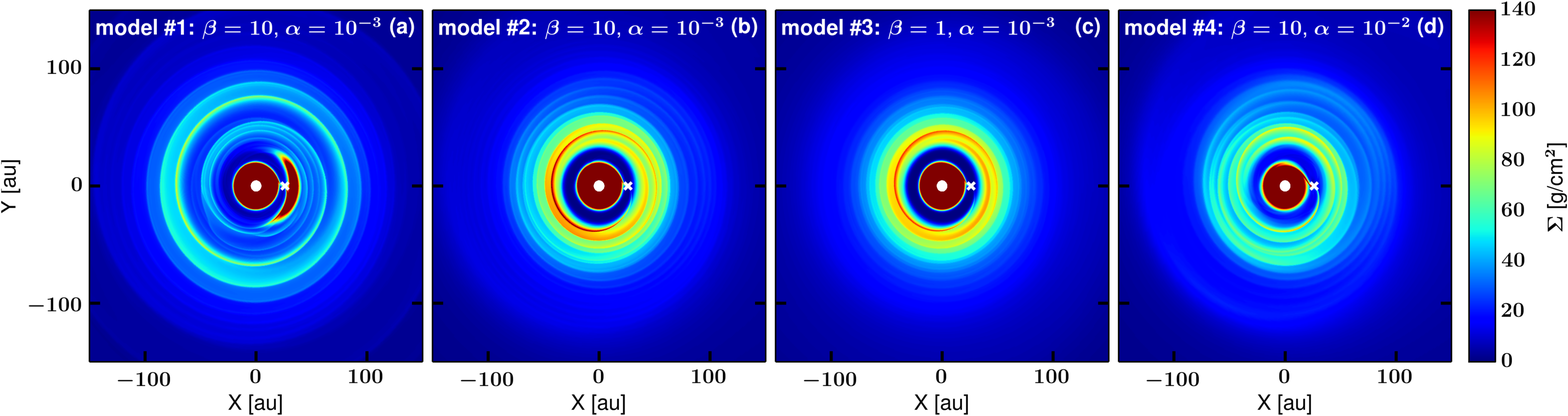}
	}
	\centerline{
		\includegraphics[width=1.0\textwidth]{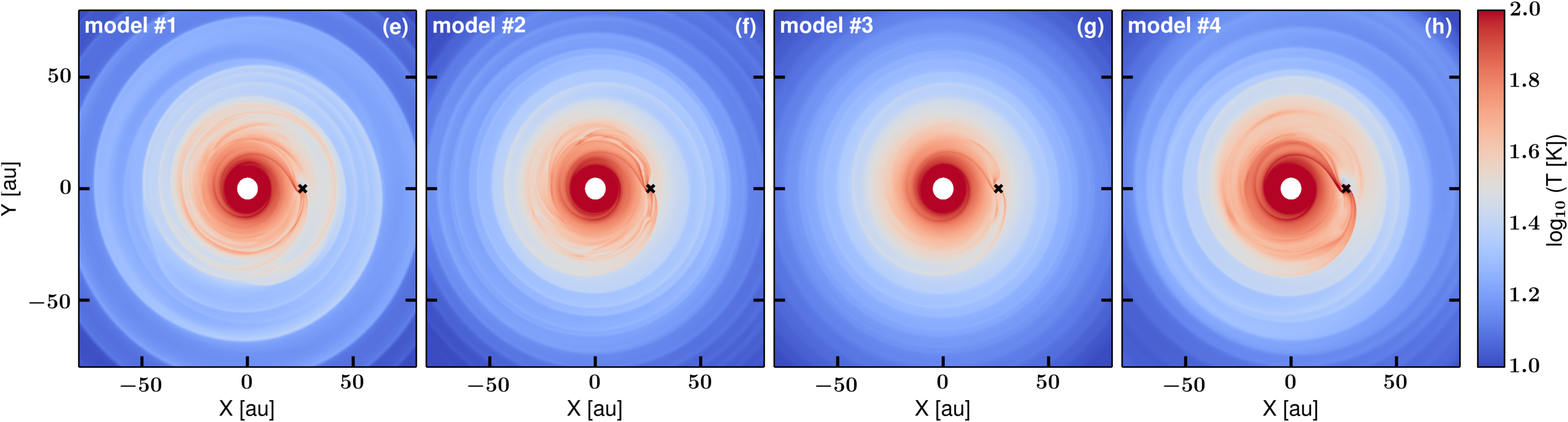}
	}	
	\caption{Surface density (top row) and temperature (bottom row) maps for a planet-to-star mass ratio of 10$^{-3}$ after 1000 planetary orbits for different cooling factors $\beta$ and viscosity values $\alpha$. The disc mass corresponds to 0.15\,M$_{\star}$. Note the difference in the x-axis scaling. \revised{Panels (\textit{a}) and (\textit{e}) show} the only model for which self-gravity is turned off. The planet's position is marked with the white and black cross, respectively.}
	\label{fig:fargo_step16}
\end{figure*}

\begin{figure*}
	\centering
	\centerline{
		\includegraphics[width=0.5\textwidth]{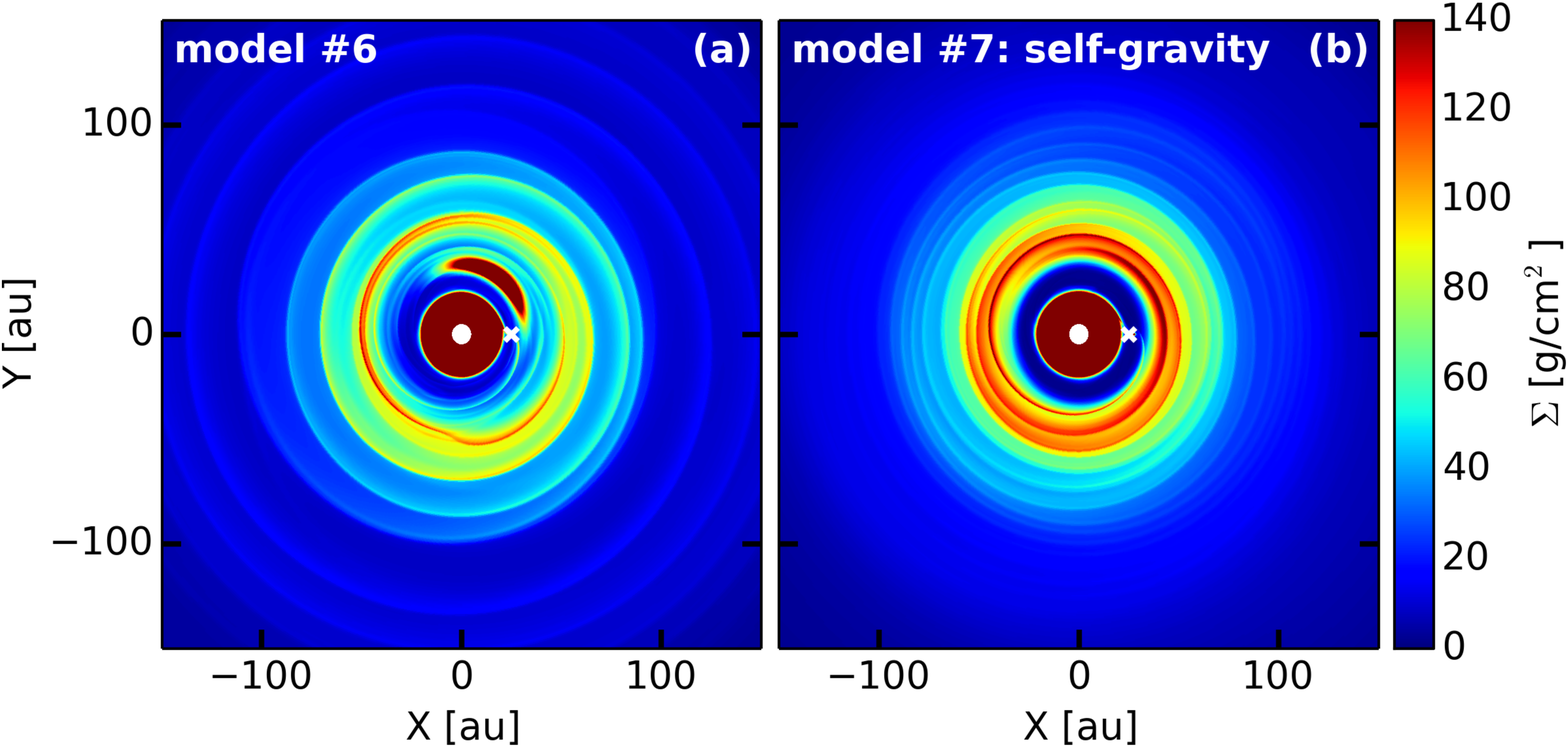}
	\hfill
		\includegraphics[width=0.5\textwidth]{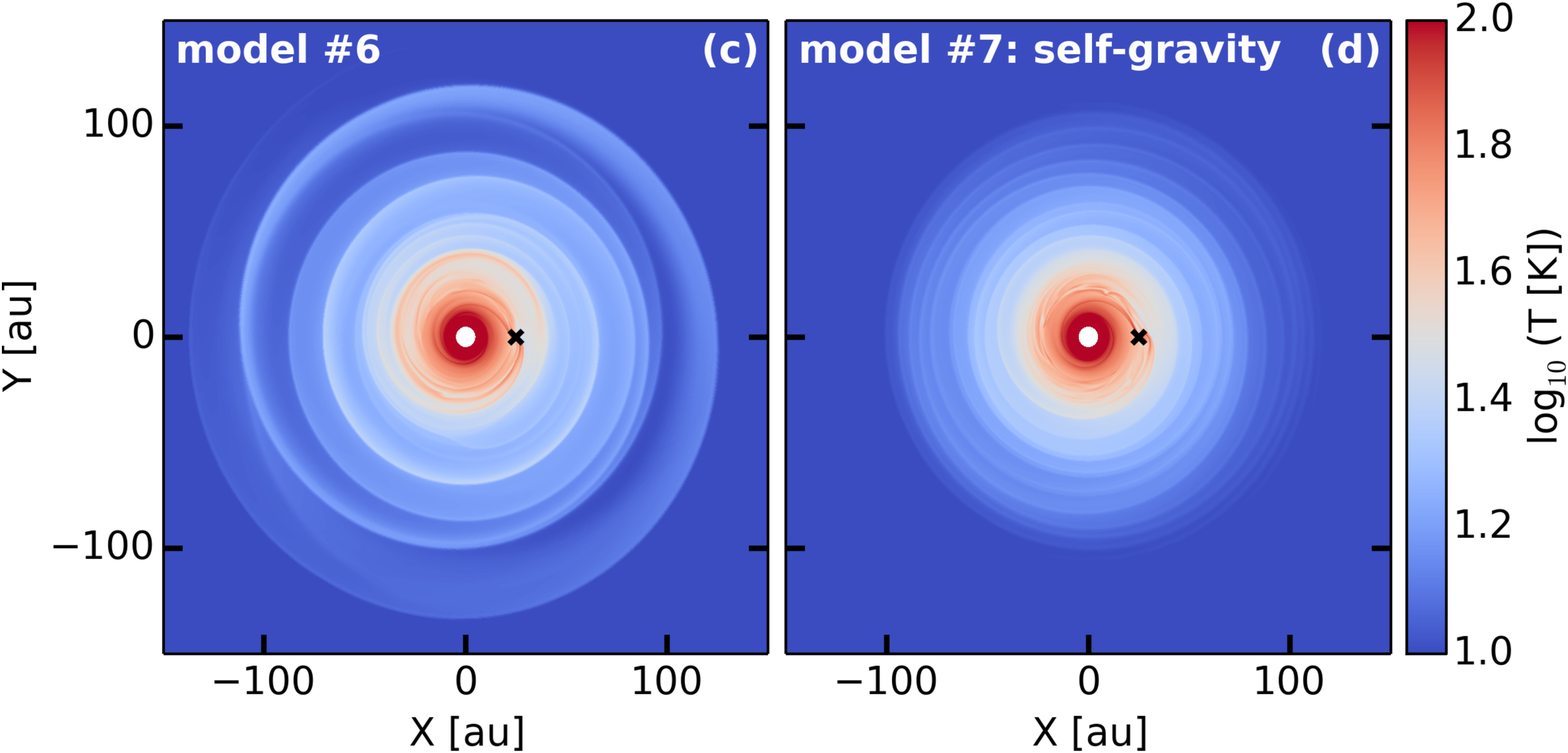}	
	}
	\caption{Surface density map for a planet-to-star mass ratio of 10$^{-3}$ after 1000 planetary orbits for a model without (\textit{a}) and with self-gravity (\textit{b}). Panels (c) and (d) show the corresponding temperature distributions. The disc mass corresponds to 0.2\,M$_{\star}$, $\beta$ is taken to be 10.}
	\label{fig:fargo_step22}
\end{figure*}

\subsection{Hydrodynamical results}
\label{subsec:results_hydro}
The numerical hydrodynamic simulations include a treatment of heating and cooling, providing a more realistic picture of the disc gas density and temperature distribution in self-gravitating discs. Hence, it is possible to study the effect of surface density and pressure scale height perturbations on the spiral morphology and contrast in scattered light images.

\begin{figure*}
	\centering
	\centerline{
		\includegraphics[width=0.5\textwidth]{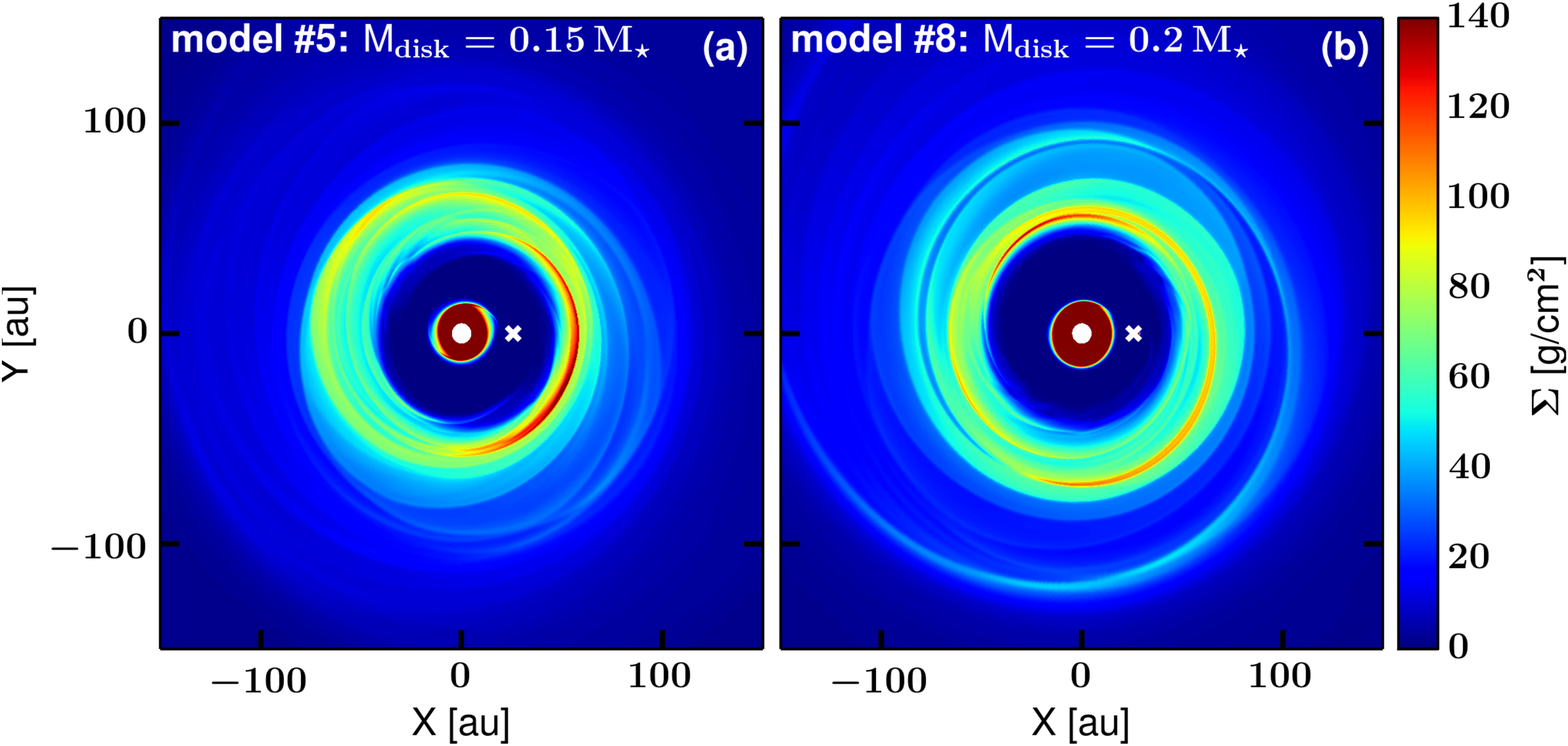}
		\hfill
		\includegraphics[width=0.5\textwidth]{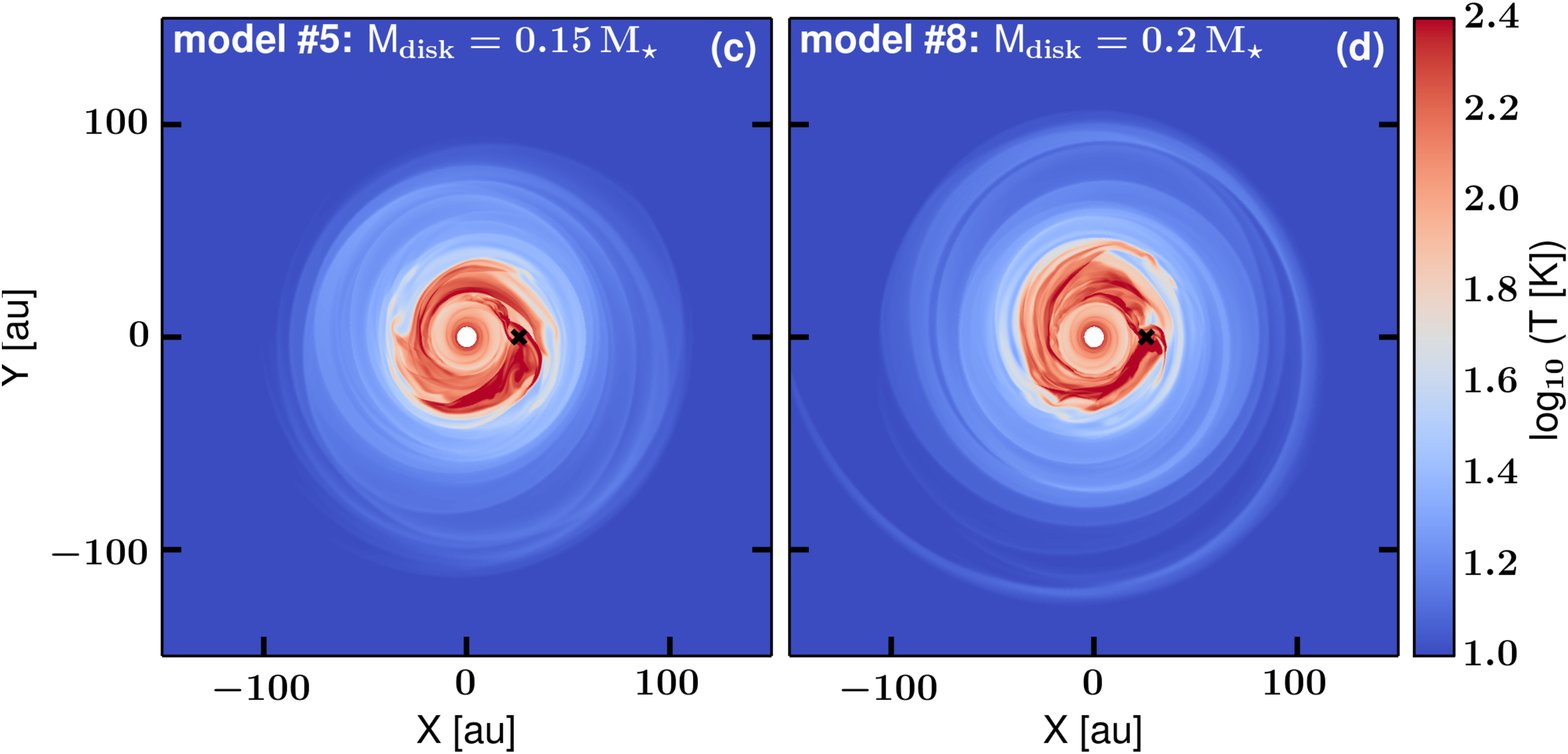}
	}
	\caption{Surface density maps (\textit{a,b}) for a planet-to-star mass ratio of 10$^{-2}$ after 1000 planetary orbits. The disc mass corresponds to 0.15\,M$_{\star}$ (\textit{a}) and 0.2\,M$_{\star}$ (\textit{b}), respectively. The temperature plots are shown in panels (\textit{c,d}).}
	\label{fig:fargo_planetmass}
\end{figure*}

\begin{figure*}
	\centering
	\centerline{
		\includegraphics[width=0.33\textwidth]{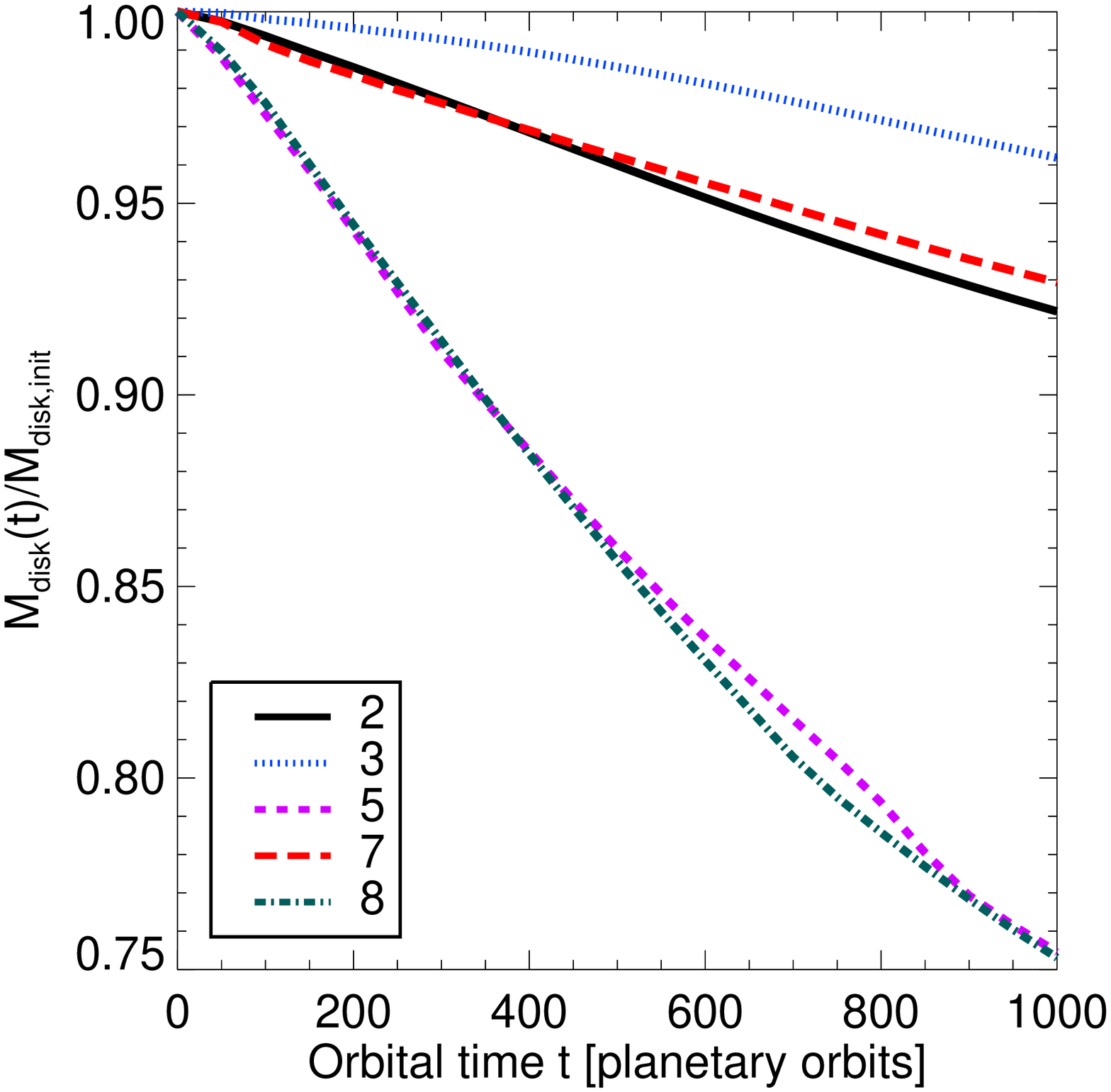}
		\hfill
		\includegraphics[width=0.33\textwidth]{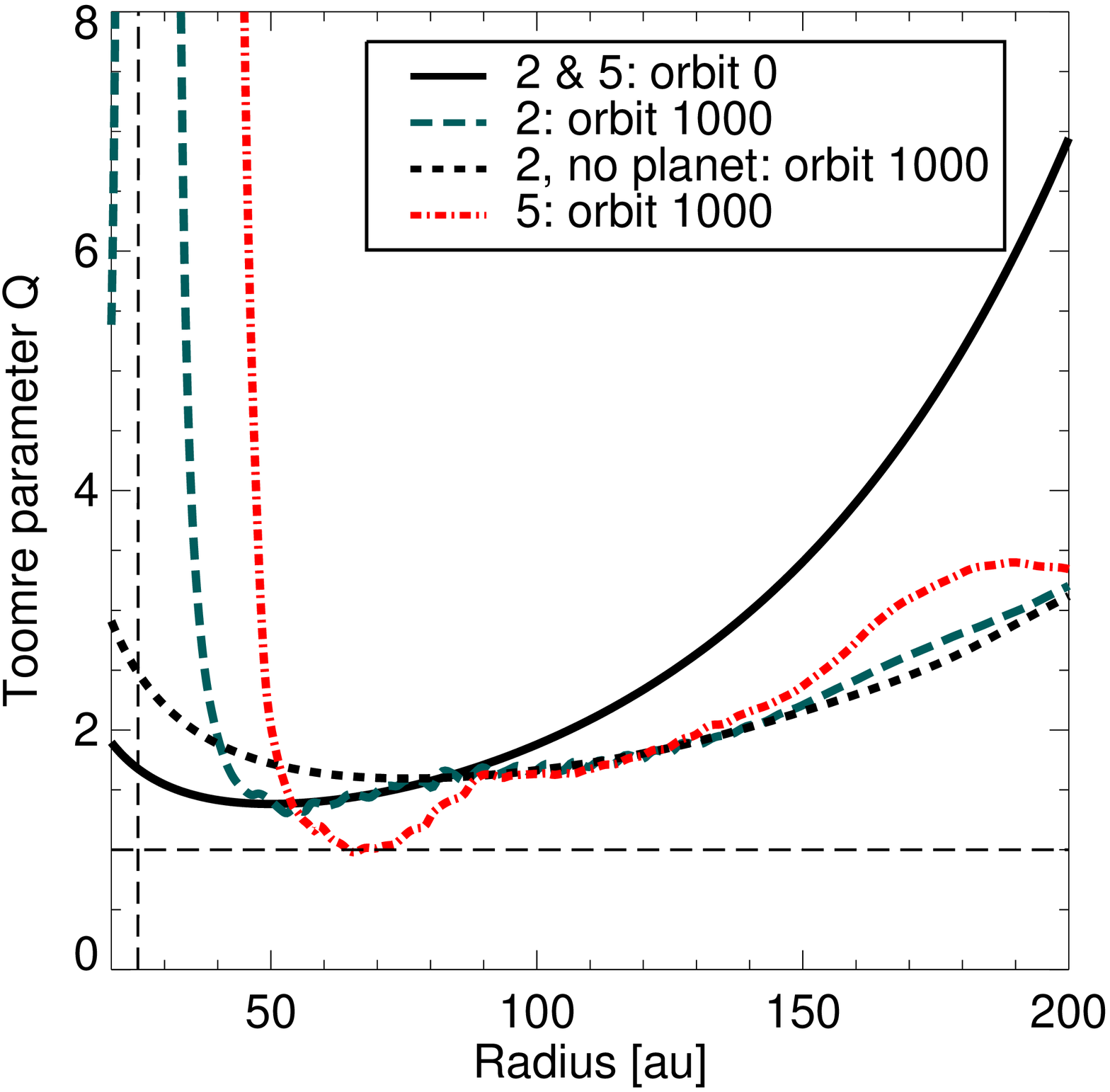}
		\hfill	
		\includegraphics[width=0.33\textwidth]{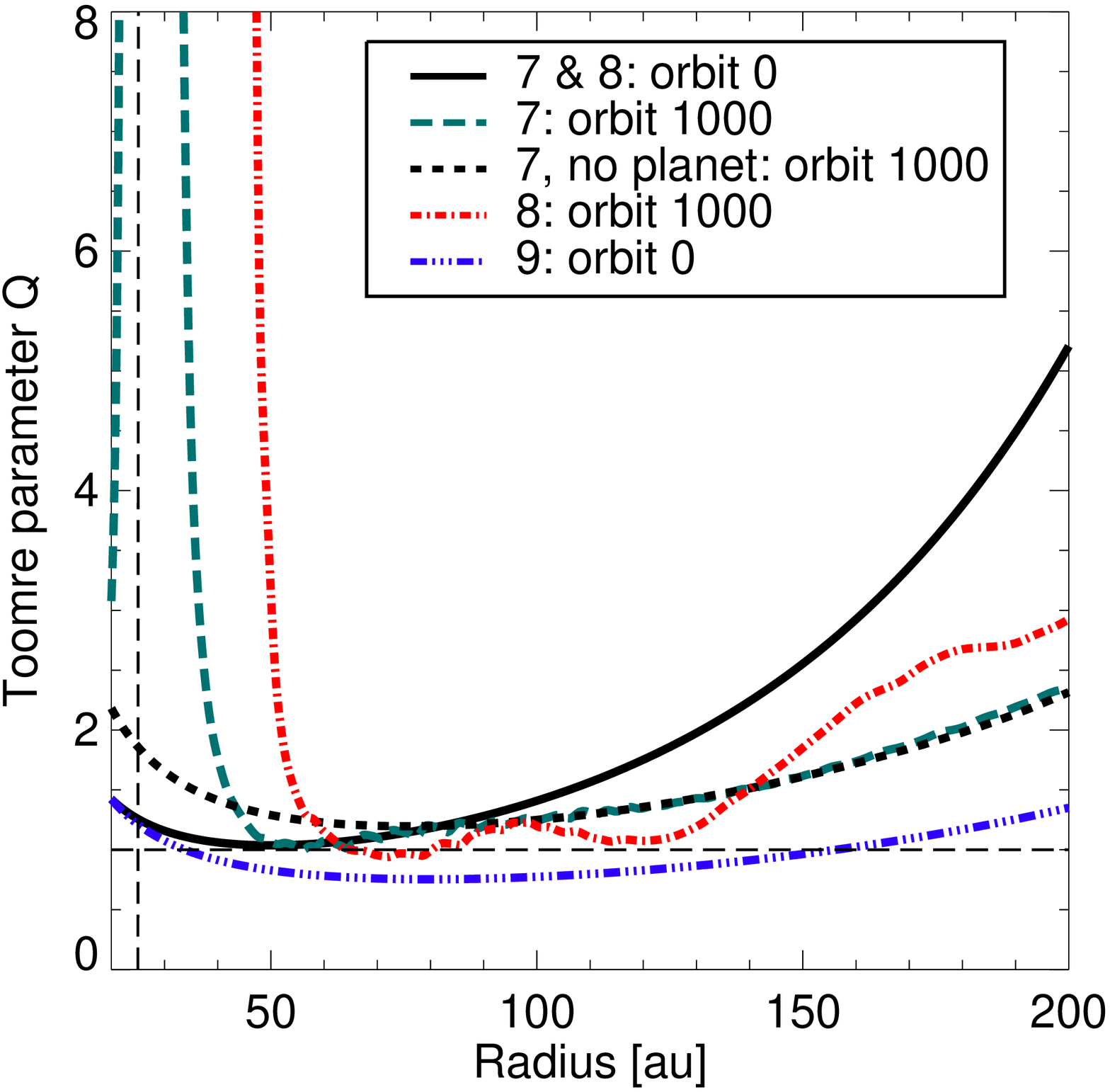}
	}
	\caption{Disc mass evolution for models with different disc and embedded planet masses (left). Azimuthally averaged Toomre parameter Q as a function of radius (middle and right panels). For radii smaller than 20 AU Q has very large values, which are not relevant for our study and therefore not plotted here. The disc mass in the middle panel corresponds to 0.15\,M$_{\star}$, for the right panel to 0.2\,M$_{\star}$. The vertical dashed line represents the position of the planet. The horizontal dashed line at Q=1 shows the critical Toomre value below which the disc is unstable to axisymmetric perturbations. A shearing disc becomes unstable to non-axisymmetric perturbations even at slightly larger values around $Q \sim 1.5-2$.}
	\label{fig:toomre_comp}
\end{figure*}

\subsubsection{Energy equation vs. locally isothermal structure}
\label{subsubsec:results_hydro_liso}
For our reference model \textit{1} the simulation considers a non self-gravitating disc with a mass of 0.15\,M$_{\star}$, a planet-to-star mass ratio of $10^{-3}$ and an $\alpha$-viscosity of $10^{-3}$. The $\beta$-cooling factor is set to 10. In Fig. \ref{fig:fargo_step16}\textit{a} a snapshot of the gas surface density after 1000 planetary orbits is shown. The embedded planet clears out a gap in less than 100 orbits. The disc viscosity is sufficiently low and the planet-to-star mass ratio is high enough, for the angular momentum flux carried by the density waves to overcome the inflow due to viscous accretion in the disc, leading to the gap opening near the planet's position. The planet is not able to produce a full cavity as seen in observations of transitional discs. This might be, however, linked to the possibility that dust located in the very inner disc close to the star could not be detected, for instance due to grain growth. Furthermore, distinct spiral density waves are induced by the planet. The dominant feature of this model is a vortex at the outer edge of the gap that moves with the local Keplerian speed and survives until the end of the simulation. The existence of this vortex is in accordance with \citet{ataiee2013}, who showed that vortices can be long-lived for moderate disc viscosity values and massive planets (cf. also \citealt{zhu2014, fu2014}). The vortex influences the surface density perturbations induced by the planet close to its position.\\

Assuming a locally isothermal equation of state corresponds to an infinitely short cooling time-scale. This means that the temperature always remains constant after a density increase since compressive work is immediately radiated away. Therefore, the decisive advantage of the energy equation simulations over locally isothermal models is that the information of the temperature distribution is tracked. In Fig. \ref{fig:fargo_step16}\textit{e-h} it can be seen that the spirals themselves are hotter than the background disc. The temperature of the spiral depends on the cooling time-scale. If the cooling time-scale is long, the disc material is unable to cool down to its pre-shock equilibrium temperature before the next heating event occurs with the passing of the next shock wave. This leads to an overall higher temperature and disc scale height, also outside of the spiral itself. Another interesting finding is that in all temperature plots, best recognizable in Fig.\ref{fig:fargo_step16}\textit{g}, the spiral close to the planet's position seems to split up in two parts. Outwards of the inner spiral and inwards of the outer spiral wake an additional high temperature arm is seen. Those structures are located within the planetary horseshoe orbit leading to a rather circular shape. They are caused by shock fronts going through the U-turn of the horseshoe orbit. We can check whether a planet can generate a strong enough perturbation in the temperature to meet the criteria for spiral detection obtained from analytical modelling by \cite{juhasz2014}. They found that a relative change of at least 0.2 in pressure scale height ($\delta H/H$) is required for the spirals to create detectable observational signatures. Calculating this relative change for our simulation with $\beta=10$ gives 0.2 at the $\varphi=270^{\circ}$ axis. This value increases to a maximum of 0.6 near the planet at a zero azimuth angle. Hence, we expect to see the spirals in scattered light (cf. Sect. \ref{subsec:results_rt}).

\subsubsection{Effect of self-gravity}
\label{subsubsec:results_hydro_sg}
In Fig. \ref{fig:fargo_step16} the gas surface density (\textit{a--d}) and temperature (\textit{e--h}) maps of four different models for a disc mass of 0.15\,M$_{\star}$ are displayed. The leftmost panel represents the reference model with a planet-to-star mass ratio of 10$^{-3}$ after 1000 planetary orbits and a cooling time of 10\,$\Omega^{-1}$. The other panels additionally include self-gravity and only differ in the $\beta$-factor and $\alpha$-viscosity value. Models \textit{2} and \textit{3} (Figs. \ref{fig:fargo_step16}\textit{b,c}) are quite different from the reference model without self-gravity. The vortex present in the latter case is smeared out very quickly due to self-gravity effects and no interaction between the vortex and density waves can occur. Additionally, the gap around the planet's orbit is much more distinct. In contrast to very low-mass planets driving a single one-armed spiral \citep{ogilvie2002}, two arms that spiral outward of the planet's position with a mutual azimuthal shift are present. Note that this m=2 mode is not a result of self-gravity, but caused by the massive planet itself. The primary spiral arm launched at the planet's position is much stronger than the one azimuthally shifted which has a quite low density contrast. Furthermore, it is noticeable that the self-gravity models show more twisted spirals with lower pitch angles. This can be explained by the fact that the disc self-gravity modifies the positions of the Lindblad resonances, between which density waves can propagate. It shifts the location of the effective resonances closer to the planet \citep{pierens2005}.\\

To study the disc stability we show the azimuthally averaged Toomre parameter Q as a function of radial distance in Fig. \ref{fig:toomre_comp}. The middle panel only includes the Toomre profiles for models with a disc mass of 0.15\,M$_{\star}$. We compare the models with an embedded planet to the case of a self-gravitating disc without a planet (solid black vs. dotted red line). It can be seen that initially Q is always larger than 1 for all models, thus the disc by itself would not become gravitationally unstable. Adding a sufficiently massive planet considerably reduces Q and puts the disc just at the limit of being gravitationally unstable between 50 and 100\,au from the star (blue dotted line). The explanation for this behaviour is the following. When a planet is present in the disc it will carve a gap around its orbit decreasing the surface density in the gap. Since most of the material from the gap is pushed away due to angular momentum exchange between the planet and the disc the surface density of the disc will increase outside of the gap edge. Hence, a sharp density jump is created. Since Q is proportional to $\Sigma^{-1}$, it is strongly increased near the planet's position within the gap, but can be reduced just behind the gap edge, where overdensities grow. Thus, if Q is low enough at the outer edge of the gap the local increase of surface density due to gap formation might tip the balance and lower Q below unity. The strong increase of Q from $\sim$ 100\,au on is caused by the exponential mass tapering and the disc mass loss (Fig. \ref{fig:toomre_comp}, left panel). The issue for a disc mass of 0.15\,M$_{\star}$ is, however, that Q is not low enough at larger radii in order to let self-gravity affect the disc dynamics in form of additional gravitational instability structures. How a larger disc mass or an increased critical taper radius affect these results is discussed in Sect. \ref{subsubsec:results_hydro_discmass}.

\subsubsection{Effect of cooling time-scale}
\label{subsubsec:results_hydro_beta}
According to Eq. \ref{eq:cooling_timescale}, for constant $\beta$, $t_{\mathrm{cool}}$ is proportional to r$^{3/2}$, so that the cooling is faster in the inner disc. This means that the temperature profile returns back to the locally isothermal background values faster in these inner parts, but in any case not faster than the local dynamical time-scale. In general, we expect to see stronger temperature changes for longer cooling times, i.e. larger $\beta$-values. The effect of the cooling time-scale can be seen by comparing models \textit{2} ($\beta=10$) and \textit{3} ($\beta=1$). A very fast cooling ($\beta \leq 1$) is very close to the locally isothermal case. In Fig. \ref{fig:fargo_step16}\textit{c,d} the final surface density and temperature show lower values in absolute strength and contrast along the spirals compared to model \textit{2} (\textit{b,f}). This slight drop in temperature also decreases the pitch angle, leading to more tightly wound spirals which are smeared out at larger radii. A larger $\beta$ value implies less effective cooling, which makes the whole disc, i.e. also the spirals, hotter. The higher the temperature, the faster the waves spread in the disc, since the sound speed is proportional to the square root of the temperature. Hence, more mass flows through the inner disc boundary for $\beta=10$ (see Fig. \ref{fig:toomre_comp}, left panel).

\subsubsection{Effect of viscosity}
\label{subsubsec:results_hydro_visc}
Another important parameter to explore is the viscosity, for which the gap profile dependence is already known (e.g. \citealt{crida2006}). If the viscosity decreases the gap becomes gradually deeper. However, the dependence of the gap characteristic on viscosity is less sensitive in numerical simulations compared to theoretical calculations. In simulations there is no simple balance between viscous, pressure and gravitational torque, since parts of the latter are transported away by density waves (\citealt{papaloizou1984,rafikov2002}). Nevertheless, by comparing our model \textit{2} ($\alpha = 10^{-3}$) with \textit{4} ($\alpha = 10^{-2}$) in Fig. \ref{fig:fargo_step16}, one can see that as viscosity increases the gap is filled with more gas. Since the primary spiral is launched at the planet's position within the gap this also influences the spiral contrast, which is consequently reduced. Furthermore, a higher $\alpha$-viscosity smears out the spiral structure and reduces the absolute surface density at the spiral position. Therefore, we use moderate values of viscosity for the rest of our models ($\alpha = 10^{-3}$).

\begin{figure}
	\centering
	\centerline{
		\includegraphics[width=0.4\textwidth]{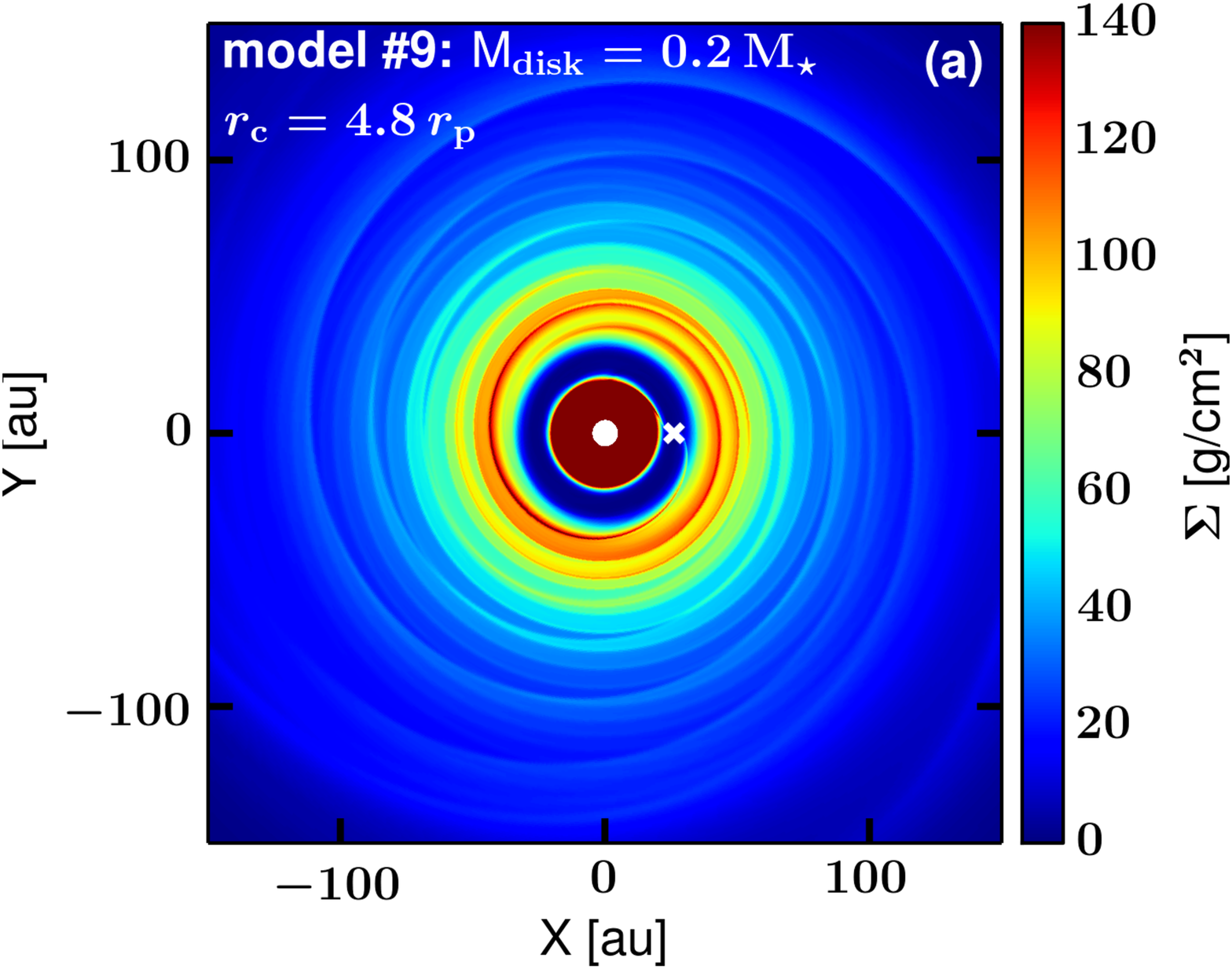}
	}
	\centerline{
		\includegraphics[width=0.4\textwidth]{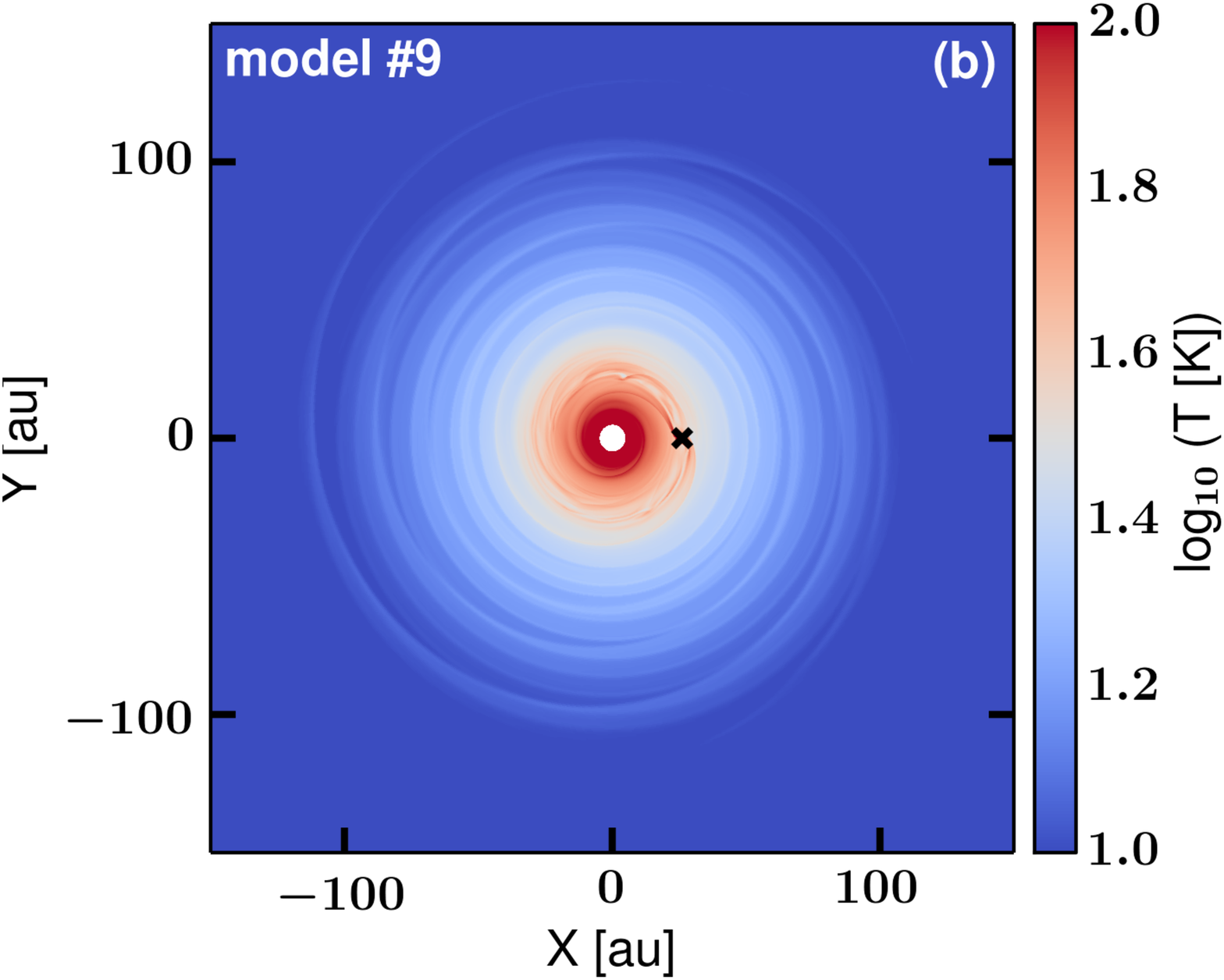}
	}
	\caption{Surface density map (\textit{a}) for a planet-to-star mass ratio of 10$^{-3}$ after 1000 planetary orbits for the model with a critical taper radius of $4.8\,r_{\mathrm{p}}$. The disc mass corresponds to 0.2\,M$_{\star}$. The temperature structure is shown in panel (\textit{b}).}
	\label{fig:fargo_rc}
\end{figure}

\begin{figure*}
	\centering
	\centerline{
		\includegraphics[width=1.0\textwidth]{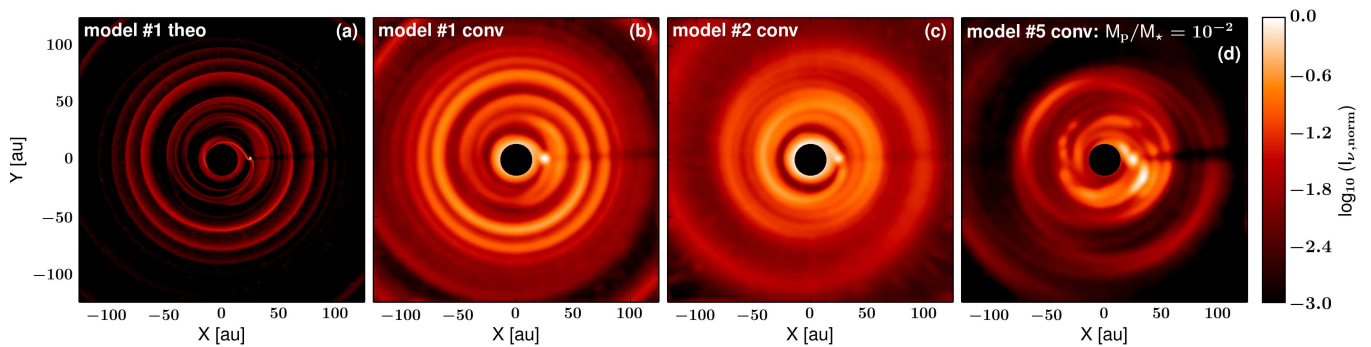}
	}
	\caption{Simulated NIR scattered light images in \textit{H}-band polarized intensity ($\lambda=1.65$\,$\mu$m). All models consider a disc mass of 0.15\,M$_{\star}$. (\textit{a}) corresponds to the reference model \textit{1} without self-gravity and shows the image at original resolution as calculated with the radiative transfer code \textsc{radmc-3d}. All other images (\textit{b-d}) are convolved with a Gaussian beam using a FWHM of 0\farcs04 (at 140\,pc distance), which is representative for observations with SPHERE/VLT in the \textit{H}-band. The central 0\farcs1 of the image were masked to mimic the effect of a coronagraph similar to real observations.}
	\label{fig:rt_step16}
\end{figure*}

\begin{figure*}
	\centering
	\centerline{
		\includegraphics[width=1.0\textwidth]{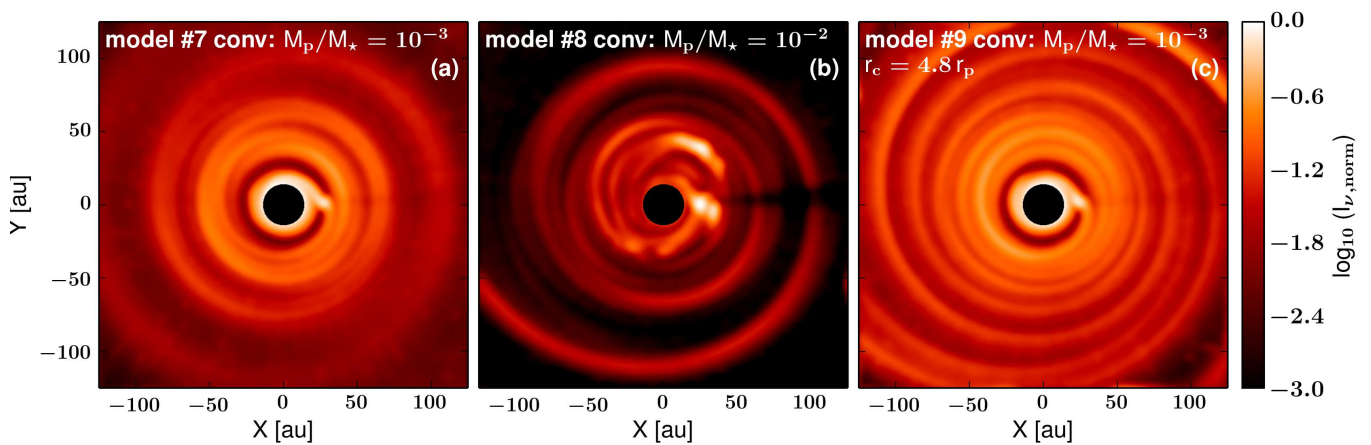}
	}
	\caption{Simulated NIR scattered light images in \textit{H}-band polarized intensity ($\lambda=1.65$\,$\mu$m). All images are based on a 0.2\,M$_{\star}$ disc. They are convolved with a Gaussian PSF with FWHM of 0\farcs04 (at 140\,pc distance). For panel (\textit{c}) the critical taper radius is set to 4.8\,r$_{\mathrm{p}}$. The central black region (0\farcs1) mimics the effect of a coronagraph similar to real observations.}
	\label{fig:rt_step22}
\end{figure*}

\subsubsection{Effects of disc and planet mass}
\label{subsubsec:results_hydro_discmass}
For the simulations \textit{6} and \textit{7} presented in Fig. \ref{fig:fargo_step22} the disc mass is increased to 0.2\,M$_{\star}$, but the planet-to-star mass ratio of $10^{-3}$ is kept. The same trends as previously described for the lower disc mass in Sect. \ref{subsubsec:results_hydro_sg} can be seen. Self-gravity (model \textit{7}) destroys the vortex and enhances the strength of the two-armed m=2 spiral considerably. This is consistent with results from \citet{lin2011}, who showed that for sufficiently
large disc mass and therefore sufficiently strong self-gravity the vortex modes are suppressed. Instead, new global spiral modes develop. Self-gravity effects become even more dominant for higher disc masses. The Toomre parameter profile for this model is plotted in the right panel of Fig. \ref{fig:toomre_comp} (solid and dashed black lines). The disc is marginally gravitationally unstable from the beginning of the simulation between the planet's position at 25\,au and $\sim$\,60\,au. This is, however, not sufficient to induce gravitational instability, since the outer disc parts remain stable. Even the perturber with a planet-to-star mass ratio of $10^{-3}$ is not capable to reduce Q to a value $\leq 1$ further out from 60\,au (dashed green line). The general effect of increasing the planet-to-star mass ratio is illustrated by comparing Fig. \ref{fig:fargo_planetmass} to Figs. \ref{fig:fargo_step16} and \ref{fig:fargo_step22}. For both disc masses the higher planet mass forces the gap to become eccentric and a distinct two-armed spiral is not seen anymore. For a $10^{-2}$\,M$_{\star}$ planet mass Q is below or close to 1 even for the outer disc region between 100 and 150\,au (red dashed line in Fig. \ref{fig:toomre_comp}). The consequence can be recognized in Fig. \ref{fig:fargo_planetmass}\textit{b}. The massive planet is able to initiate gravitational instability, visible as a quite open spiral arm in the outer disc. Particular attention has to be paid to the temperature structure. The massive planet produces significant shock heating within the inner 50\,au of the disc. These temperature changes cause perturbations in the vertical structure of the disc and following Eq. \ref{eq:scale_height} this strongly influences the characteristics of our NIR scattered light images (cf. Sect. \ref{subsec:results_rt}).\\

Apart from changing the planet mass the other possibility to trigger gravitational instability in the outer disc is to change the distribution of mass as a function of radius by increasing the mass in the outer disc for a given disc mass. This is done in our simulation with 0.2\,M$_{\star}$ by shifting the critical mass taper radius outwards to 4.8\,r$_{\mathrm{p}}$ (=120\,au). The corresponding surface density and temperature plots are illustrated in Fig. \ref{fig:fargo_rc}\textit{a,b}. A more open, global spiral pattern is seen. These large scale spirals reflect the onset of gravitational instability, but close to the planet they overlap with the planetary density waves. This instability is supported by the Toomre parameter profile in Fig. \ref{fig:toomre_comp}, right panel (blue dashed-dotted line). The disc is initially gravitationally unstable in most locations.\\

When the disc is at the limit of being gravitationally unstable only a slight increase in disc mass changes the non-axisymmetric structures considerably. This effect can be seen in Fig. \ref{fig:fargo_planetmass}. For a disc mass of 0.2\,M$_{\star}$ (Fig. \ref{fig:fargo_planetmass}\textit{b}) a spiral arm in the outer disc induced by gravitational instability is present. Its pitch angle is significantly higher than for the planetary induced spirals from models with a lower disc mass (Fig. \ref{fig:fargo_planetmass}\textit{a}).

\subsection{Synthetic radiative transfer images}
\label{subsec:results_rt}
We aim to investigate the characteristics of the main observational features in transition discs, i.e. gaps and spiral structures, for which full three-dimensional radiative transfer simulations are required. We are interested in the number of spiral arms, their pitch angle, and the brightness contrast between the spiral features and the surrounding background disc to investigate the observability of spirals. We expect that the shock heating along the spiral may increase the scale height of the disc locally. This causes bumps on the disc surface, which are irradiated by the star, and thus produce a significant brightness contrast. Based on the model setup described in Sect. \ref{sec:rt_modeling} scattered light images in polarized intensity at 1.65\,$\mu$m are calculated. The disc is nearly face-on with an inclination angle of $10^{\circ}$. The wavelength at which the following images are made is chosen such that the images are directly comparable to \textit{H}-band observations made with SPHERE/VLT and HiCIAO/Subaru. For a realistic comparison to observations the radiative transfer images are first convolved with a 0\farcs04 Gaussian beam, assuming the source to be at 140\,pc. Then, each pixel intensity value is multiplied with $r^2$, where r is its distance from the star, in order to compensate for the falloff of the stellar irradiation. This way the outer disc structures become much better visible. A coronagraph is mimicked by masking the inner 0\farcs1 of the disc (14\,au at 140\,pc distance). The images are normalized to the highest disc surface brightness while using a dynamic range of 1000 for plotting the images in logarithmic scale.

\subsubsection{Images from models without self-gravity}
\label{subsubsec:results_rt_nosg}

\begin{figure}
	\centering
	\centerline{
		\includegraphics[width=0.5\textwidth]{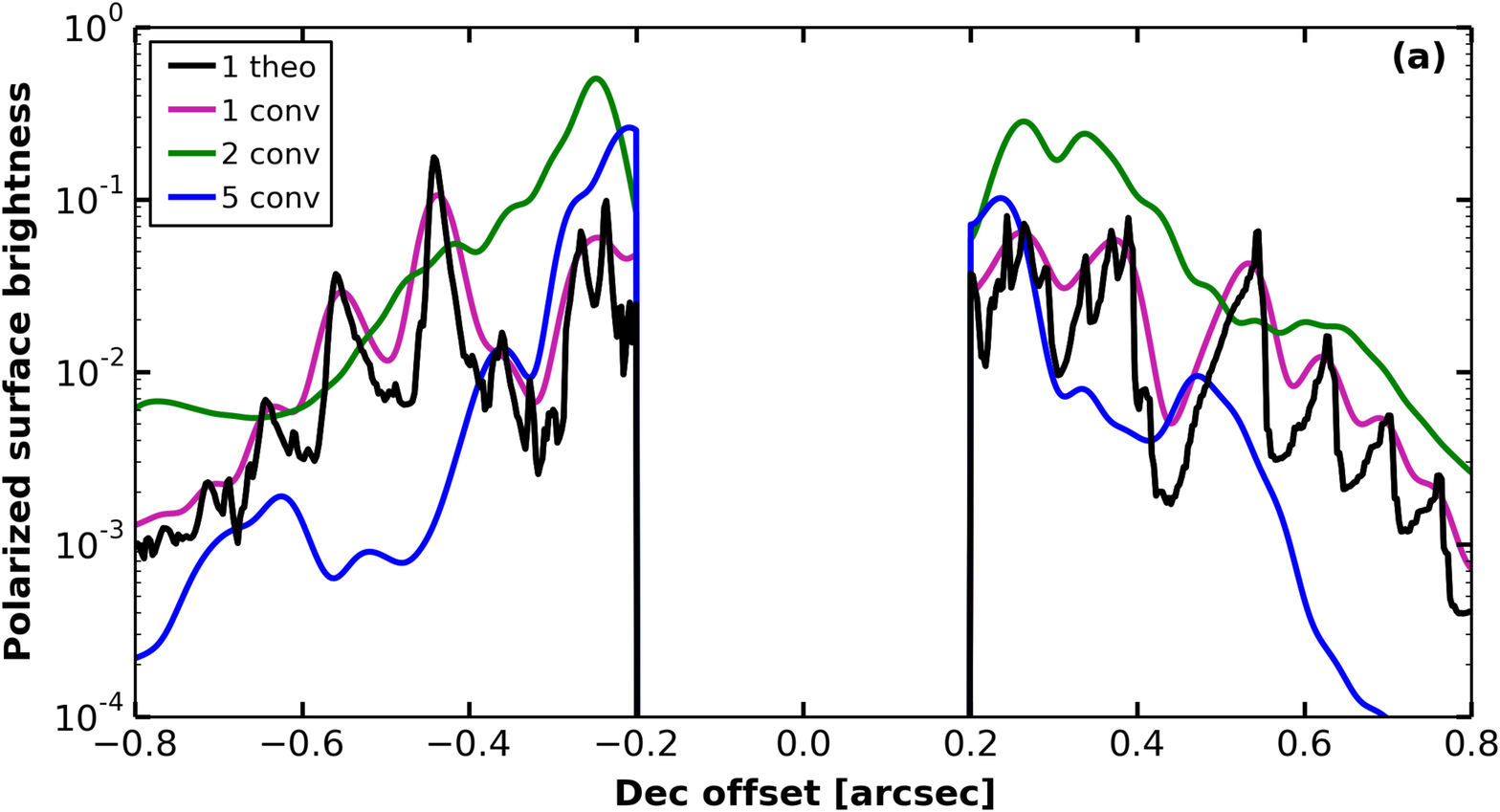}
	}
	\centerline{
		\includegraphics[width=0.5\textwidth]{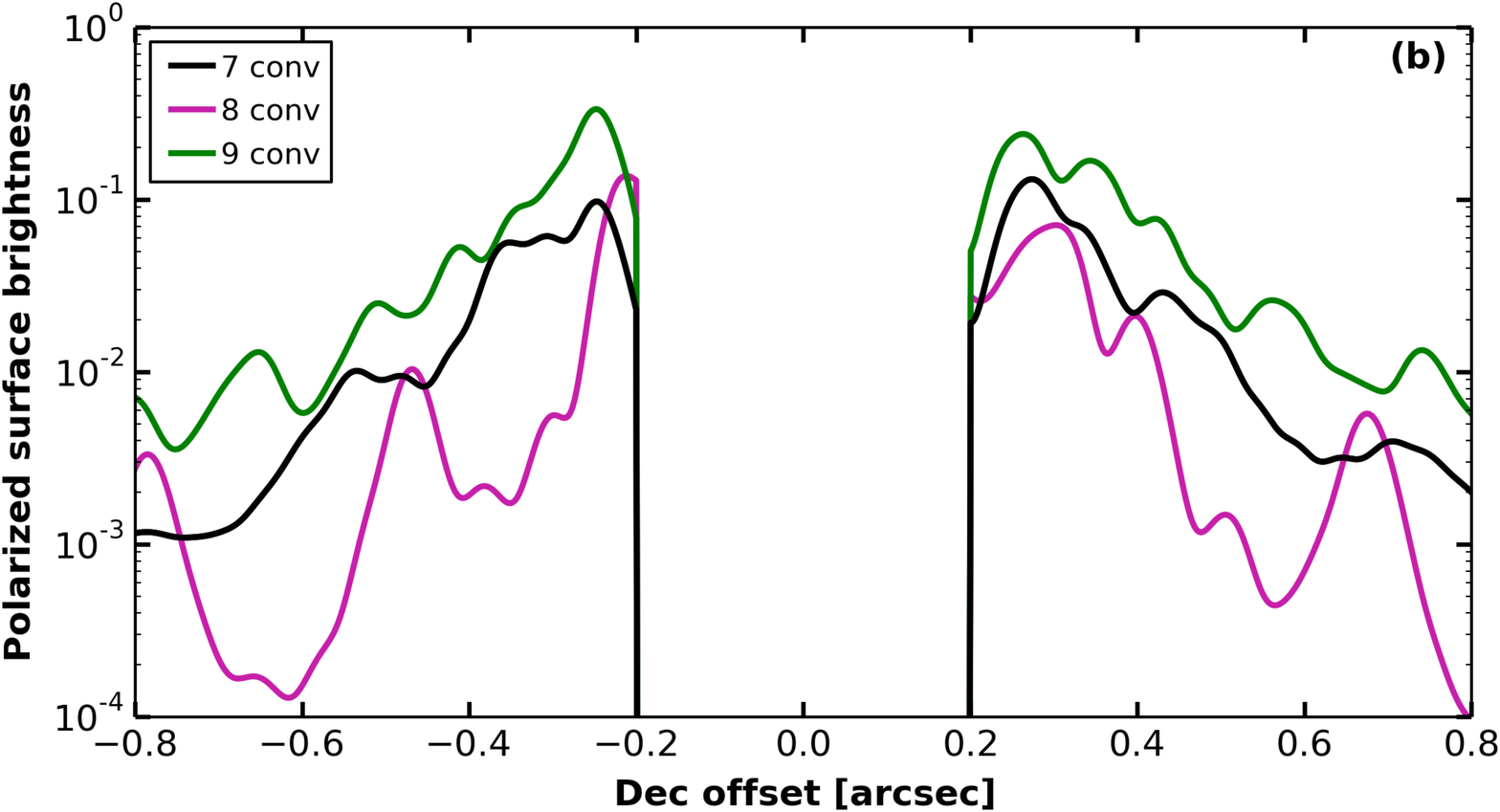}
	}	
	\caption{Radial profiles of the normalized polarized surface brightness belonging to the calculated images in Figs. \ref{fig:rt_step16} and \ref{fig:rt_step22}, respectively. \revised{Since the focus is on measuring the contrast of the outer spirals, the profiles are only plotted beyond an inner circle with a radius of 0\farcs2.}}
	\label{fig:brightness}
\end{figure}

The planetary m=1 spiral is clearly visible in the scattered light image at infinite resolution based on the reference model \textit{1} without self-gravity (see Fig. \ref{fig:rt_step16}\textit{a}). The brightness contrast between the spiral and the contiguous disc ranges from 12 to 30. The large vortex present in the hydrodynamical simulations is seen in the scattered light as well. This has, however, to be interpreted with caution, since it is related to the two-dimensional character of our hydrodynamical simulations. Vortices can extend through the whole disc from the midplane to the atmosphere (cf. \citealt{zhu2014}), but can be only accurately simulated with three-dimensional hydrodynamical models. If the vertical elongation of the vortex is small, it is expected to appear less extended in scattered light, which traces the disc surface layer. After convolving this theoretical image with the Gaussian 0\farcs04 PSF the vortex as well as the planet-induced spirals are significantly smeared out, but still clearly visible. As can be seen in Fig. \ref{fig:brightness}\textit{a}, the contrast of the spirals with the surrounding disc is reduced by a factor of about three, giving contrast values between 3 and 10 (magenta line) for a resolution of 0\farcs04. \revised{We note that the contrast is expected to be further reduced in observations due to instrument related and additional noise effects.}

\subsubsection{Images from models including self-gravity}
\label{subsubsec:results_rt_sg}
In all simulations shown in Figs. \ref{fig:rt_step16}\textit{c,d} and \ref{fig:rt_step22}\textit{a-c} self-gravitating discs are considered. First, we study the effect of self-gravity by comparing the results of models \textit{1} and \textit{2}. As already explained in Sect. \ref{subsubsec:results_hydro_sg}, the inclusion of self-gravity destroys the vortex immediately. This leads to a more distinct primary spiral launched close to the planet's position (see Fig. \ref{fig:rt_step16}\textit{c}). Furthermore, the radial brightness cross-sections in Fig. \ref{fig:brightness}\textit{a} illustrate that the clear wiggle structure from the models without self-gravity (magenta line) is smeared out (green line). Self-gravity forces the spiral close to the planet to be more tightly wound. This in turn reduces the total number of individual spiral windings seen in the scattered light, since they cannot be distinguished anymore. The contrast of the spiral with the surrounding disc varies between 1.5 to 10 for a resolution of 0\farcs04.\\

The effect of increasing the planet-to-star mass ratio is displayed in Figs. \ref{fig:rt_step16}\textit{d} and \ref{fig:rt_step22}\textit{b}. The absolute brightness near the planet is increased, leading to several bright spots. Moreover, the gap is not fully detectable anymore in the scattered light. The irregular structure close to the planet's position is caused by the perturbation in temperature (cf. Fig. \ref{fig:fargo_planetmass}\textit{c,d}). Thus, this heating strongly increases the disc scale height within the gap. This elevated region is now irradiated by the star hiding the gap in the scattered light image. At the same time structures characteristic of gravitational instability are visible. They occur as quite open spirals reaching also outer disc parts. The spiral contrast in the outer disc is extremely high with values of up to 40. A second possibility to create gravitational instability structures is a model with a disc that is gravitationally unstable from the beginning, i.e. no massive planet has to work as a trigger. The resulting scattered light image in Fig. \ref{fig:rt_step22}\textit{c} shows a quite symmetric spiral for the whole disc. The planet-induced spiral is only visible close to the planetary orbit and the outer disc is dominated by the spirals triggered by gravitational instability. For this model the polarized surface brightness in Fig. \ref{fig:brightness}\textit{b} has a clear wiggle structure (green line) with a brightness contrast of approximately 3. This contrast is similar to the images from models without self-gravity, but a factor of 10 smaller than for a self-gravitating disc model with a massive planet ($10^{-2}$\,M$\star$). This shows that the brightness contrast between the spirals and the contiguous disc is strongly influenced by the planet-to-star mass ratio in combination with the disc mass distribution. As soon as gravitational instability spirals triggered by a planet develop, the contrast increases tremendously, the spiral shape is strongly influenced and its pitch angle slightly increases.\\

Another feature to mention, which is present in all images, is the shadow of the planet along a position angle of $\mathrm{PA}=-90$\,deg. This is just an artefact of our simulation methods, since we do two-dimensional hydrodynamical, but three-dimensional radiative transfer simulations. This means that a surface density enhancement, e.g. a circumplanetary ring, will increase the total density at all disc heights in the radiative transfer models. It is unlikely that even a circumplanetary disc around a massive planet with several tens of M$_{\mathrm{jup}}$ can cast a shadow over the whole outer disc in its full vertical extent. The shadow is expected to be seen in the midplane, but the disc atmosphere should remain nearly unaffected.

\section{Discussion and Summary}
\label{sec:summary}

We present hydrodynamical simulations of planet-disc interactions considering viscous heating, cooling, and additionally self-gravity effects. Subsequently, the density and temperature structure resulting from these simulations are used for the follow-up radiative transfer modelling, in order to predict synthetic scattered light images from a protoplanetary disc around a typical Herbig Ae star. The focus is set on analysing the morphology and detectability of the main observational features of transition discs in scattered light observations, distinct gaps, and non-axisymmetric spiral arms.\\

The conclusions of this paper are as follows:

\begin{enumerate}[leftmargin=*,label={\arabic*.}]
	\item{} A clear m=2 spiral structure, and in some cases even higher azimuthal wave number modes, can be excited by a single planet in a self-gravitating disc. The shape of the planet-induced spirals depends on the planet mass and disc parameters, such as the viscosity, and heating/cooling time-scales.\\
	
	\item{} When self-gravity of the disc is taken into account the spirals are found to have a smaller pitch angle, making them more tightly wound. For sufficiently large disc mass, and therefore sufficiently strong self-gravity, the vortex modes are suppressed, and the strength of the spiral is enhanced considerably.\\
	
	\item{} All our calculated polarized intensity images show a planetary gap and spiral arms with a variety of morphologies depending on planet mass, disc mass distribution, and the influence of the disc's self-gravity. Convolving the images with a circular Gaussian PSF with a FWHM of 0\farcs04 (typical for current 8-10m class telescopes) lowers the contrast by a factor of $\sim 3$, but the spiral features are still visible.\\

	\item{} The scale height along the spiral increases due to the temperature changes, forming local bumps on the disc surface, and thus creates a signal in the reflected light. Therefore, the contrast in the images also depends on the cooling time-scale via the $\beta$-factor. In a disc where the cooling is very efficient (i.e $\beta \lesssim 1$), which approximates the case of a locally isothermal disc, any temperature increase is immediately suppressed. The remaining density perturbation alone is not capable of producing a sufficiently strong spiral contrast in scattered light to be observable with current state-of-the-art telescopes (cf. \citealt{juhasz2014}).\\
	
	\item  Our result of a non-self-gravitating disc with an embedded planet shows that the combination of planet-induced density and scale height perturbation along the spiral due to shock heating is sufficient to achieve the contrast between the spiral and the contiguous disc necessary to be detectable in observations. The contrast values estimated for a resolution of 0\farcs04 are in a range from 3 to 10, which is consistent with the brightness contrast of spirals seen in recent observations.\\
	
	\item{} There are two models in our series for which the Toomre parameter profiles allow the creation of spirals due to gravitational instability. However, their initiation mechanism is different in both cases. For model \textit{9} the disc itself is massive enough, and the mass distribution allows it to be gravitationally unstable from the beginning of the simulation. Spirals are visible in the whole disc, a difference is, however, seen in the pitch angle between the spirals close to the planet and in the outer disc. The spiral is more tightly wound near the planet's position as it is driven by the planet, while the spirals in the outer disc are created by gravitational instability showing a larger pitch angle. A long-scale spiral with quite regular windings is visible in the corresponding scattered light image with a contrast of about 3 for a resolution of 0\farcs04.\\
	
	\item{} The second possibility is a disc that is initially at the edge of being gravitationally unstable. As soon as a sufficiently massive planet is included, gravitational instability can be triggered by this planet (model \textit{8}). A spiral arm with a huge brightness contrast of $\sim 40$ and with a higher pitch angle than the spirals driven by a planet in a gravitationally stable disc is generated. The strong temperature perturbations near the high-mass planet partly hide the low-density gap and dominate the irregular structures in this disc region.
\end{enumerate}

To summarize, perturbations in the vertical structure of the disc, i.e. scale height perturbations caused by temperature variations due to planet-induced accretion heating or local heating by gravitational instability, are able to create the necessary spiral-background contrast seen in scattered light observations. The presence of a sufficiently massive planet embedded in a marginally gravitationally stable disc can lead to a variety of spiral morphologies depending on the planet and disc mass. It is possible to create a tightly-wound spiral close to the planet and a more open spiral arm in the outer disc. However, the explanation of the origin of symmetric, open double-armed spirals seen e.g. in SAO 206462 remains challenging. \revised{Recently, three-dimensional global hydrodynamical simulations (isothermal and adiabatic, assuming a non self-gravitating disc) combined with radiative transfer calculations after a short-time of evolution (10--20 local orbits) have shown that inner spirals caused by massive planets (6~M$_{\mathrm{jup}}$ at $\sim$\,100\,au) can be visible at scattered light with similar pitch angle, extension and symmetry as observational results (\citealt{dong2015}). However, at longer times of evolution when the disk reaches a quasi-steady state it is expected that such a planet opens a visible gap in the outer disk (100--200\,au) contrary to observational results at NIR and (sub-)mm.} With our model setup there is either one dominant primary spiral \revised{in the outer disc} when the planet is working as a trigger for gravitational instability, or higher mode spirals with similar contrasts are seen when the disc is already initially unstable. A second planet would be required in order to create two primary spirals of a m=1 or m=2 structure with nearly the same contrast. \revised{However, for this scenario the two planets would need to have just the right mass ratios and be located exactly at the right radial locations and azimuthal angles in the disc to cause symmetric spirals.} In addition, other mechanisms apart from self-gravity may play a role for non-axisymmetric structure formation as well. We note that a caveat of our approach is that we do not perform fully consistent radiation hydrodynamics simulations, e.g. including the consistent release of accretion energy from the planet via radiation, since this is quite computational expensive for a parameter study. \revised{Long-term simulations in three dimensions that include these effects are needed to investigate the differences with our current results and for the direct comparison with observations. This will be addressed in future work.} Additionally, we are aware that the cooling law from \citet{gammie2001}, used also in most of the previous work about self-gravitating discs, is a simplified description. It has to be elaborated in order to calculate the disc temperature structure more precisely, which influences the local scale height changes, and thus also the brightness contrast in scattered light.

\section*{Acknowledgments}
We are grateful to C.~Dominik for useful discussions and remarks on the manuscript. We thank C.~Baruteau for his support in solving some numerical problems with the \textsc{fargo} code. A.~P. is a member of the International Max Planck Research School for Astronomy and Cosmic Physics at the University of Heidelberg, IMPRS-HD, Germany. A.~P. acknowledges the CPU time for running simulations on bwGRiD, member of the German D-Grid initiative, funded by the Bundesministerium f\"ur Bildung und Forschung and the Ministerium f\"ur Wissenschaft, Forschung und Kunst Baden-W\"urttemberg. P.~P. is supported by Koninklijke Nederlandse Akademie van Wetenschappen (KNAW) professor prize to Ewine van Dishoeck. A.~J. acknowledges support of the DISCSIM project, grant agreement 341137 funded by the European Research Council under ERC-2013-ADG.


\bibliographystyle{mnras}
\bibliography{mnras_pohl}



\appendix


\bsp	
\label{lastpage}
\end{document}